\documentclass[notitlepage,12pt]{jedm}
\usepackage[table]{xcolor}
\usepackage{url}
\usepackage{hyperref}
\usepackage{graphicx}
\usepackage{subcaption} 
\usepackage{amsmath} 
\usepackage{amsfonts} 
\usepackage[normalem]{ulem}
\useunder{\uline}{\ul}{}
\usepackage{rotating}
\usepackage{pdflscape}

\hypersetup{
  colorlinks   = true, 
  urlcolor     = blue, 
  linkcolor    = blue, 
  citecolor   = blue 
}

\begin{document}

\title{Predictive and explanatory models might miss informative features in educational data}
\date{} 

\author{{\large Nicholas T. Young}\\Michigan State University\\youngn18@msu.edu \and {\large Marcos. D Caballero}\\Michigan State University\\University of Oslo\\caball14@msu.edu}

\maketitle

\begin{abstract}
We encounter variables with little variation often in educational data mining (EDM) due to the demographics of higher education and the questions we ask. Yet, little work has examined how to analyze such data. Therefore, we conducted a simulation study using logistic regression, penalized regression, and random forest. We systematically varied the fraction of positive outcomes, feature imbalances, and odds ratios. We find the algorithms treat features with the same odds ratios differently based on the features' imbalance and the outcome imbalance. While none of the algorithms fully solved how to handle imbalanced data, penalized approaches such as Firth and Log-F reduced the difference between the built-in odds ratio and value determined by the algorithm. Our results suggest that EDM studies might contain false negatives when determining which variables are related to an outcome. We then apply our findings to a graduate admissions data set. We end by proposing recommendations that researchers should consider penalized regression for data sets on the order of hundreds of cases and should include more context about their data in publications such as the outcome and feature imbalances.
\\ 

{\parindent0pt
\textbf{Keywords:} random forest, penalized regression, feature imbalance, outcome imbalance
}
\end{abstract}

\section{Introduction}
When working with educational data, we often encounter imbalanced binary input and outcome features, by which we mean the variable is not equally split into its two categories. For example, demographics in science, technology, engineering, and mathematics (STEM) are often imbalanced due to historical and ongoing injustices. While data mining with imbalanced data has been studied extensively \cite{chawla_data_2009}, less attention has been paid to the types of imbalanced data that appear in educational data mining (EDM) studies.

For example, educational data sets might consist of a single course on the order of a hundred students (and hence a hundred data points) to the entire university or even multiple universities, resulting in hundreds of thousands of data points. Furthermore, educational data often includes continuous, categorical, and binary variables. As a result, an educational data set might contain many features with different imbalances as well as an imbalanced outcome. For specific examples of these occurring in EDM studies, we refer the reader to the following papers \cite{arreola_bayesian_2020,chiu_gender_2020,kruzicevic_predictors_2012,manduca_improving_2017,marquez-vera_predicting_2013,pelaez_using_2019,rovira_data-driven_2017,sax_women_2016,spoon_random_2016,tai_examining_2017,young_identifying_2019}.

In the context of logistic regression, which is a popular EDM technique \cite{pena-ayala_educational_2014} and was a common technique used in the previously cited studies, much work has focused around outcome imbalance and how to work with such data. When the outcome is imbalanced, the regression coefficients and the probabilities generated from the logistic regression model are biased \cite{king_logistic_2001}. To correct for these biases, various techniques such as Rare Events Logistic Regression \cite{king_logistic_2001}, Firth penalized regression \cite{firth_bias_1993}, and introducing a log-F distributed penalty \cite{greenland_penalization_2015} have been proposed, which we explain in depth in Sec.~\ref{sec:background}.

More recently, machine learning techniques have become popular in educational research. One example is random forest \cite{pena-ayala_educational_2014,breiman_random_2001}. Just as logistic regression has biases that might be relevant to EDM data, random forest is also known to have such biases. In particular, random forest ranks categorical features with many levels \cite{strobl_bias_2007} and continuous features higher than categorical features with fewer levels \cite{nicodemus_letter_2011} when determining which features are most predictive of an outcome.

Most interesting for the context of this paper is a study by \citeN{boulesteix_random_2012} building on the work of \citeN{nicodemus_letter_2011}. In their paper, they systematically varied the amount of predictive information that each binary feature contained as well as the feature imbalance and then used random forest as well as a variant better suited for categorical features, conditional inference forest \cite{strobl_bias_2007}, to compare how well the algorithms could detect the informative features from the noise. Their key finding was that features with higher imbalances were ranked lower than features with lower imbalances even when they had the same ``built-in'' amount of predictive information. A later study \cite{nembrini_revival_2018} extended the work by including continuous features as well as binary features but only examined the case when none of the features contained predictive information. These studies suggest that when modeling data, our results might be measuring spurious properties of the features, i.e. their imbalance, rather than their predictive information.

In this study, we seek to extend this line of work by considering the data typical of EDM studies. That is, data that includes a mix of continuous, categorical, and binary features with varying degrees of predictive or explanatory ability, and a binary outcome feature that might be imbalanced. In addition, new techniques for ranking random forest features have been developed, such as the AUC-permutation importance \cite{janitza_auc-based_2013}, which are designed for imbalanced data sets and hence, might prove fruitful for EDM research. Finally, we wish to extend the work to regression techniques commonly used for educational data and explore how these biases might manifest in these techniques.

Specifically, we ask three research questions:
\begin{enumerate}
    \item How might known random forest feature selection biases change when the outcome is imbalanced as is often the case in EDM studies, and does the AUC-permutation importance affect those biases?
    \item How might known machine learning biases manifest in traditionally explanatory techniques such as logistic regression?
    \item How might penalized regression techniques successfully applied in other disciplines be used in EDM to combat any discovered biases?
\end{enumerate}

For the purposes of this paper and our research questions, we will define \textit{imbalance} as the degree to which a binary feature deviates from the case of being split equally between categories and \textit{bias} as a systematic preference by a model toward one feature over another given both features have the same level of association with the outcome or toward one feature over another given the first feature has a smaller association with the outcome compared to the second feature. Our interest in this paper is then when imbalance in a feature results in a bias in the model and hence, would be problematic for a researcher or practitioner. For example, in their concluding remarks \citeN{boulesteix_random_2012} notes that a bias caused by imbalance may be viewed as a "serious problem" if one assumes the features ranked highly by an importance measure have the strongest association with the outcome rather than category frequencies that provide the highest value to an importance measure. We adopt this view in this paper. As \citeN{ding_theoretical_2019} notes, we make certain commitments to our analysis when we decide on a particular approach to quantification and our questions. Our commitments might disagree if we have different approaches or questions.

For this paper, it should also be noted that our overarching goal is to compare existing approaches to analyzing data typically found in EDM research and not to introduce our own new promising method for analyzing such data.

The rest of the paper proceeds as follows. In Sec.~\ref{sec:background}, we provide an overview of the algorithms and approaches we mentioned in the introduction and that we use in the rest of the paper. In Sec.~\ref{sec:methodology}, we explain how we constructed our simulation data and carried out our neutral comparison simulation study \cite{boulesteix_plea_2013}. In Sec.~\ref{sec:results}, we provide the results of our simulation study. In Sec.~\ref{sec:realdata}, we apply what we learned in the simulation study to a graduate admissions data set from United States universities. In Sec.~\ref{sec:discussion}, we provide answers to our research questions, compare our findings with similar studies, and consider how our choices might have influenced the results. In Sec.~\ref{sec:future_work}, we propose future directions for this work, both in terms of the data and algorithms. Finally, in Sec.~\ref{sec:conclusion}, we provide the conclusions from our study and outline a set of recommendations.

\section{Background}\label{sec:background}
Here, we introduce the two paradigms of statistical modeling and then provide an overview of the algorithms we used in our study.

\subsection{Paradigms of Statistical Modeling}
When discussing modeling data, there are two prominent paradigms and goals, both of which are used in EDM: prediction and explanation/information \cite{breiman_statistical_2001,romero_educational_2020,aiken_framework_2021}. \citeN{shmueli_explain_2010} provides an overview of these approaches and we summarize the key points here.

Explanatory modeling or explanation is focused on the causal effect of some set of inputs $X$ on some outcome $Y$. That is, given some data set, explanation is concerned with which inputs produce a statistically significant effect when modeling the outcome. Traditional logistic or linear regression are examples of explanatory models. Under this approach, models are evaluated based on how well they fit the data using some statistic. In the case of logistic regression or linear regression, common statistics are Pseudo-$R^2$ and $R^2$.

In contrast, prediction is focused on generating a model for analyzing new data and determining the outcome and not necessarily the causal effect. Under this paradigm, having two sets of data, one to train the model and one to test the predictive capabilities of the model, is essential as to provide an estimate of the model's predictive ability.

Because prediction is not focused on the causal effects, statistical significance has no role in assessing features in predictive models. Instead, features are assessed based on whether they improve predictions of the model. While a feature with a small effect might be statistically significant, it might not have predictive power because a predictive model might perform just as well without the feature as with it \cite{shmueli_explain_2010,greenberg_predictive_1997}.

As a corollary to this, we should not expect a model with high explanatory power to necessarily have high predictive power or vice versa, and hence, features with high explanatory power might not have high predictive power. \citeN{shmueli_explain_2010} includes an example where the model with the highest explanatory power does not have the highest predictive power and conditions under which that may occur.

While our motivation and research questions start with predictive modeling and then move to explanatory modeling, we will introduce explanatory methods first because the logistic regression background is needed for the penalized regression section of the predictive methods.

\subsection{Methods we use as Explanatory}
\subsubsection{Traditional Logistic Regression}
When the outcome, $Y$, is binary, logistic regression is the standard technique for explanatory modeling. Under this approach, the probability, $p$, of finding the outcome of $Y=1$ is given by

\begin{equation}
    \log_b\frac{p}{1-p}=\beta_0+\beta_1 x_1+\beta_2 x_2 +...+ \beta_n x_n
\end{equation}

when $x_1, x_2,..,x_n$ are the input features and the $\beta$ are the coefficients. Under this formula, logistic regression has a similar form as linear regression.

We can rearrange the equation to solve for the odds which becomes
\begin{equation}
    odds(x_1,x_2,...,x_n)=\frac{p}{1-p}=b^{(\beta_0+\beta_1 x_1+\beta_2 x_2 +...+ \beta_n x_n)}
\end{equation}

where $b$ is traditionally the natural base, $e$.

Under this formulation, it makes sense to talk about the odds ratio (OR) or the change in odds as a result of increasing an input feature $x_j$ by 1 unit. More formally,

\begin{equation}
    OR_{x_j}=\frac{odds(x_1,x_2,...x_j+1,..,x_n)}{odds(x_1,x_2,...x_j,..,x_n)}=e^{\beta_j}
\end{equation}

which means that the exponentials of the coefficients correspond to the odds ratio for each feature. Notice that the odds ratio is independent of the value of $x_j$.

Because a $\beta$ of 0 means no effect, an odds ratio of 1 is equivalent to no effect \cite{theobald_beyond_2019}. Likewise, an odds ratio greater than 1 means an increase in the odds while an odds ratio less than 1 means a decrease in the odds.

An important caveat to this is what a unit increase is and what the odds ratio is in reference to. Often, continuous features are normalized so that the mean is 0 and the variance is 1 or scaled so that an increase of a unit has a tangible meaning. For example, SAT scores are only reported in multiples of 10 so scoring one point higher on the SAT is meaningless. Instead, the researcher would want to adjust the scale of the scores so that an increase of 1 unit corresponded to 10 points better on the test (or another meaningful increment).

For continuous features, what the odds ratio is in reference to is answered by the scale choice. For categorical features, especially unordered categorical features, the answer is nontrivial. An increase of 1 unit might not be meaningful or even possible (e.g., what would an increase of 1 unit of race mean?). In that case, it is customary to use one-hot encoding and create separate, binary features for each label. For example, for race, we could create 6 features: white, Asian, Black, Latinx, Native, Multi-racial. Under this approach with binary features, an increase of a unit corresponds to changing categories, such as Black compared to non-Black student, which depends on the arbitrary choice of which label is assigned $x_j=1$ and which is assigned $x_j=0$. As \cite{theobald_beyond_2019} notes, it is often preferable to invert the odds ratios which are less than 1 to easily compare all odds ratios, which is equivalent to swapping our label for $x_j=0$ and $x_j=1$.

\subsubsection{Penalized Regression}
When the data contains issues that might make modeling difficult (i.e., small sample size, correlations, and more features than data points), adding a penalty to logistic regression might be beneficial. This idea is based on the bias-variance trade-off in which we can increase the bias of the coefficient to reduce its variability or vice versa \cite{hastie_elements_2009}. As a result, penalized regression can be useful for feature selection, which is often an important first step in EDM \cite{pena-ayala_educational_2014}.

For typical least squares regression with $m$ features and $n$ cases, we are trying to solve the expression

\begin{equation}
    \text{argmin}_{\beta}(||Y- X^T \beta||^2)
\end{equation}

where $Y$ is a $n \times 1$ vector of the outputs, $\beta$ is a vector of the $m \times 1$ vector coefficients, and $X$ is a $m \times n$ matrix of the input data.

When we use penalized regression instead, we add a penalty, $\mathbf{P}$, that might depend on the coefficients or data.

\begin{equation}
    \text{argmin}_{\beta}(\;||Y-X^T \beta ||^2+\mathbf{P}(\beta,X))
\end{equation}

In this study, we consider two types of penalization for explanatory methods, Firth and Log-F penalization, although many more exist. See \citeN{ensoy_separation_2015} for an overview of methods often used in cases of separation, where an input feature perfectly predicts the outcome, or rare events.

Under Firth penalization, we try to combat the asymptotic bias of the coefficient estimates, which inversely depend on the sample size to some power. Specifically, the Firth method adds a penalty that removes the asymptotic bias to order $\mathcal{O}(n^{-1})$, making it especially useful for small data sets \cite{firth_bias_1993}. It does so by penalizing the Jeffreys invariant prior \cite{jeffreys_invariant_1946}, which is inversely related to the amount of information in the data. That is, the penalty is larger the less the data allows us to determine the coefficients. For a simple one-feature model, the penalty is equivalent to adding 0.5 to each cell of the 2x2 contingency table of the feature and the outcome \cite{heinze_solution_2002}, making  this penalization especially useful in the case of separation. In theory, this penalization should then shrink the confidence intervals of the features with more imbalance because more uncertainty would have resulted in a higher penalty.

The Jeffreys invariant prior is not without issues, such as being dependent on the data, which are summarized in \citeN{greenland_penalization_2015}. To overcome these, Greenland and Mansournia proposed a $\text{log-F}(m,m)$ distributed penalty. The penalty has a tuning parameter, m, that controls the amount of penalization with a higher $m$ providing more accurate estimates of smaller $\beta$ but less accurate estimates of larger $\beta$. When little is known about the data, Greenland and Mansournia recommend taking $m=1$ to allow for a wider range of possible values. For a single parameter model, the choice $m=1$ makes the Log-F penalty equivalent to the Firth penalty.

In addition to overcoming issues with the Jeffreys prior, the log-F penalty can be implemented via data augmentation, meaning that any software capable of performing logistic regression can also do Log-F penalization. For a chosen $m$, the researcher adds $m$ pairs of rows to their data for each feature, where one row has outcome $Y=1$ and the other has outcome $Y=0$. In the pair of rows, the researcher then selects one feature to have value 1 and all of the other features to have values of 0, with the choice of feature unique to each pair of rows. The weights for each row are set to be $m/2$ and any intercept feature should be set to 0 in these added rows. An example of this for a 2-feature model with $m=1$ is shown in Table \ref{tab:logf_table}.

\begin{table}[]
\centering
\caption{Log-F data augmentation example for a two feature and m=1 example. The last four rows are the augmented data.}
\label{tab:logf_table}
\begin{tabular}{ccccc}
\hline
Outcome & Feature 1 & Feature 2 & Intercept & Weight \\ \hline
1       & 0.748     & 0.10      & 1         & 1      \\
...     & ...       & ...       & ...       & ...    \\
1       & 1         & 0         & 0         & 1/2    \\
0       & 1         & 0         & 0         & 1/2    \\
1       & 0         & 1         & 0         & 1/2    \\
0       & 0         & 1         & 0         & 1/2    \\ \hline
\end{tabular}
\end{table}

It should be noted that despite similarity in name, log-F penalized regression has no relation to the recently proposed LogCF framework \cite{chen_logcf_2020}.

\subsection{Methods we use as Predictive}
\subsubsection{Penalized Regression}
In addition to using penalized logistic regression as an explanatory method, there are also penalties designed for using regression as a predictive tool. Two of the most common are ridge and lasso, which are described in detail in \citeN{hastie_elements_2009}.

Ridge penalization adds a penalty to the regression equation proportional to the square of the $\beta$s.

\begin{equation}
    \text{argmin}_{\beta}(\;||Y-X^T \beta||^2+\lambda ||\beta||^2)
\end{equation}

Equivalently, it requires the sum of the squared $\beta$ coefficients to be less than some value.

\begin{align}
    \text{argmin}_{\beta}(\;||Y-X^T \beta||^2) \\
    \text{subject to} \sum_{j=1}^m \beta_j^2 \leq t
\end{align}

Here, $\lambda$, or equivalently $t$, controls the degree of penalization, with a higher value associated with a stronger penalty.

Ridge penalization is often used in cases of multi-collinearity because it reduces the variability of the coefficients. That is, for two correlated features without penalization, one could be extremely positive and the other extremely negative to offset each other. With the squaring of the coefficients under ridge penalization, the coefficients can no longer offset each other and hence, must shrink. Mathematically, ridge penalization is equivalent to scaling each $\beta$ by $\frac{1}{1+\lambda}$.

Instead of penalizing based on the squared $\beta$, we can penalize based on the absolute value of the $\beta$; this is the premise of lasso penalization. Mathematically, lasso penalization seeks to solve

\begin{equation}
    \text{argmin}_{\beta}(\;||Y- X^T \beta||^2+\lambda |\beta|)
\end{equation}

Equivalently, it requires the sum of the absolute value of the $\beta$ coefficients to be less than some value.

\begin{align}
    \text{argmin}_{\beta}(\;||Y-X^T \beta||^2) \\
    \text{subject to} \sum_{j=1}^m |\beta_j| \leq t
\end{align}

Again, $\lambda$ controls the amount of penalization. Here though, the lasso penalty is designed for feature selection because it shrinks some $\beta$ to zero while shifting the values of the others.

Lasso is not designed for correlated features, and hence, it can encounter issues in those cases. For example, if two features are correlated, either could be shrunk to zero without reducing the accuracy of the model. Therefore, lasso can exhibit variability concerns under correlation.

One way around this is to combine the penalties into a single penalty, which is the idea between elastic net \cite{zou_regularization_2005}. Mathematically, the elastic net penalty is

\begin{equation}
    \text{argmin}_{\beta}(\;||Y- X^T \beta||^2+\lambda (\alpha ||\beta||^2+(1-\alpha)|\beta|))
\end{equation}
.

where $\lambda$ controls the overall penalization and $\alpha$ controls the amount of mixing of the lasso and ridge penalties, with the special case $\alpha=0$ reducing to lasso penalization and $\alpha=1$ reducing to ridge regression.

While these algorithms are typically used for prediction, various methods for using these algorithms in an explanatory manner have been developed along with corresponding p-values or other feature selection techniques \cite{cule_significance_2011,hofner_controlling_2015,meinshausen_stability_2010,lee_exact_2016,lockhart_significance_2014}. We will only use these algorithms as predictive tools, but we include references to these approaches here for completeness.

\subsubsection{Forest Methods}
Random forest is an ensemble method of decision trees based on the Classification and Regression Trees (CART) framework \cite{breiman_random_2001}. For each decision tree, a subset of features, often noted $mtry$, is randomly selected and used to predict the outcome. To grow the tree, features are split into two groups with the specifics of the splits determined by which ones minimize the Gini Index, a measure of variance, the most. After all trees have been grown, the algorithm uses some method of aggregating results, such as a majority vote of the trees, to determine what the overall prediction is. Because the features are split, categorical features do not need to be one-hot encoded like they would in logistic regression.

To determine which features are relevant to the prediction, the features are often assessed by the mean decrease in the Gini Index across all trees, with a larger value meaning the feature is more predictive of the outcome. However, \citeN{strobl_bias_2007} showed that the Gini Index is biased toward continuous features and features with many categories. That is, because continuous and features with many categories have many possible split points, it more likely the algorithm can find an ideal split than the algorithm could for a binary feature that has only 1 split point. Therefore, these features will be viewed as more important because they appear to better separate the classes.

As a result, alternative measures such as accuracy permutation importance have become popular. To use accuracy permutation importance, each feature is randomly permuted one at a time and the change in predictive accuracy is recorded. The idea is that when a feature that is more predictive of the outcome is permuted, the predictive accuracy will decrease more than when a feature with less predictive information is permuted. As a result, the changes in predictive accuracy can be used to rank the features in the model qualitatively. More recently, an alternative based on the AUC, which is the probability that the positive case ranks higher than the negative case over all possible pairs of positive and negative cases, has been proposed by \citeN{janitza_auc-based_2013}. This AUC-permutation importance is claimed to perform better than the accuracy permutation importance measure when the outcome is imbalanced. It is important to note that both these importances only make sense in the context of the model and relative to each other.

Because the Gini Index is also used to create feature splits, the entire algorithm can be biased when the data contains binary, categorical, and continuous features (as is often the case in EDM). To correct this problem, \citeN{strobl_bias_2007} proposed conditional inference forests, which are based on the conditional inference framework \cite{hothorn_unbiased_2006}. Rather than minimize the Gini Index to find ideal splits, conditional inference forests use the conditional inference independence test to determine which feature to split and how to split it. Simulation studies by \citeN{strobl_bias_2007} have found that using conditional inference forests with subsampling without replacement does in fact, correct the biases shown by traditional random forest.

For more details about these algorithms, see the supplemental material of \citeN{young_identifying_2019}.

\section{Methodology}\label{sec:methodology}
In this section, we describe how we conducted our simulation study, including creating the synthetic data and modeling the data.

For the simulation study, we seek to understand how different variants of random forest and random forest importance measures rank features with identical odds ratios but different imbalances. We also seek to understand how accurately different variants of logistic regression can determine the built-in odds ratio, whether the features with built-in odds ratios different from 1 are determined to be statistically significant, and how those answers vary with feature and outcome imbalance. Based on studies described in the introduction and background, we expect that features with higher imbalance should rank lower than features with lower imbalance for a given odds ratio for random forest algorithms. We also expect that logistic regression algorithms should be more accurate in determining the underlying odds ratio for feature with lower imbalance than features with higher imbalance for a given odds ratio.

To do so, we apply four random forest and importance measure pairs and six traditional and penalized logistic regression algorithms to fifteen data sets, representing three sample sizes and five levels of outcome imbalance. We created these data sets to be representative of data that an EDM researcher might encounter or have been described in the literature.

Readers interested in learning more about the advantages and disadvantages of the algorithms we used based on the literature and the results of this study are encouraged to look ahead to Table \ref{tab:algorithm_pros_cons}.

\subsection{Data Creation}
To conduct our simulation study, we first needed to generate our simulated data. To create binary features with varying degrees of imbalance and information, we considered a 2x2 contingency table, Table \ref{tab:cont_table}. We used labeling conventions similar to those of \citeN{olivier_effect_2013} for the reader's convenience because we reference their formulas here.

\begin{table}[b]
\caption{2x2 contingency table of fractions for generic binary feature.}
\label{tab:cont_table}
\centering
\begin{tabular}{cccc}
\hline
      & $x_j=0$         & $x_j=1$         & Total \\ \hline
$Y=0$   & $\pi_{00}$    & $\pi_{01}$    & $\pi_{0+}$      \\
$Y=1$   & $\pi_{10}$    & $\pi_{11}$    & $\pi_{1+}$      \\
Total & $\pi_{+0}$    & $\pi_{+1}$    & 1.0  \\ \hline
\end{tabular}
\end{table}

For some binary feature $x_j$, let the fraction of cases with $x_j=0$ be $\pi_{+0}$ and the fraction of cases with $x_j=1$ be $\pi_{+1}$. Likewise, for the binary outcome feature $Y$, let the fraction of cases with $Y=0$ be $\pi_{0+}$ and the fraction of cases with $Y=1$ be $\pi_{1+}$. Then the feature imbalance is represented by the ratio $\pi_{+0}:\pi_{+1}$ and the outcome imbalance is represented by $\pi_{0+}:\pi_{1+}$. We pick $x_j=1$ and $Y=1$ to be the minority classes, though the choice is arbitrary.

To quantify the amount of information contained in a feature for predicting or explaining the outcome, we will use the odds ratio which is $OR=\frac{\pi_{00} / \pi_{01}}{\pi_{10} / \pi_{11}}=\frac{\pi_{00}\pi_{11}}{\pi_{10}\pi_{01}}$ using the notation in Table \ref{tab:cont_table}, where an odds ratio further from 1 signifies more information.

By specifying the feature imbalance (in the form of $\pi_{+1}$), the outcome imbalance (in the form of $\pi_{1+}$) and the odds ratio, we can uniquely express the values in the 2x2 table. Furthermore, any one of the three can be changed while the remaining two can be held constant, allowing us to manipulate the feature imbalance, the outcome imbalance, and the odds ratio systematically. An example with fractions for a hypothetical data set is shown in Table \ref{tab:example_data}, where (b) changes the feature imbalance from .4 to .2, resulting in a more imbalanced feature, (c) changes the outcome imbalance from .5 to .25 resulting in a more imbalance outcome, and (d) changes the odds ratio from 1 to 2.79, resulting in a feature with more information about the outcome.

\begin{table}[t]
    \caption{Examples of changing only one of the feature imbalance, outcome imbalance, or odds ratio.}
    \label{tab:example_data}
    \begin{subtable}{.5\linewidth}
      \centering
        \caption{Reference table}
        \begin{tabular}{cccc}
              & $x_j=0$         & $x_j=1$         & Total \\ \hline
        $Y=0$   & .30    &  .20   &   .50   \\
        $Y=1$   &  .30  &  .20   &  .50    \\
        Total &  .60  & .40   & 1  \\ \hline
        \end{tabular}
    \end{subtable}%
    \begin{subtable}{.5\linewidth}
      \centering
        \caption{Changing only the feature imbalance}
        \begin{tabular}{cccc}
              & $x_j=0$         & $x_j=1$         & Total \\ \hline
        $Y=0$   & .40    & .10   &   .50   \\
        $Y=1$   & .40   &  .10   &   .50   \\
        Total & .80   &  .20  & 1  \\ \hline
        \end{tabular}
    \end{subtable}
        \begin{subtable}{.5\linewidth}
      \centering
        \caption{Changing only the outcome imbalance}
        \begin{tabular}{cccc}
              & $x_j=0$         & $x_j=1$         & Total \\ \hline
        $Y=0$   & .45    &  .30   &    .75  \\
        $Y=1$   & .15   &  .10   &   .25  \\
        Total &  .60  & .40   & 1  \\ \hline
        \end{tabular}
    \end{subtable}%
    \begin{subtable}{.5\linewidth}
      \centering
        \caption{Changing only the odds ratio}
        \begin{tabular}{cccc}
              & $x_j=0$         & $x_j=1$         & Total \\ \hline
        $Y=0$   & .36    &  .14   &   .50   \\
        $Y=1$   & .24  &  .26   &   .50   \\
        Total &  .60  & .40  & 1 \\ \hline
        \end{tabular}
    \end{subtable}
\end{table}

To determine the values in the 2x2 table, we can rearrange the formula for the odds ratio in terms of $\pi_{+1}$, $\pi_{1+}$, and $\pi_{11}$ found in the literature to solve for $\pi_{11}$ \cite{olivier_effect_2013}. Doing so, we find that

\begin{equation}
    \pi_{11}=\frac{1+(\pi_{+1}+\pi_{1+})(OR-1)-Q}{2(Q-1)}
\end{equation}
where

\begin{equation}
    Q=\sqrt{(1+(\pi_{1+}+\pi_{+1})(OR-1))^2+4OR(1-OR)\pi_{+1}\pi_{1+}}
\end{equation}

In the case that $OR=1$, that is the feature contains no predictive or explanatory information for the outcome, the expression for $\pi_{11}$ is indeterminate. In that case, the feature and outcome are independent so $\pi_{11}=\pi_{+1}\pi_{1+}$.

Once we know $\pi_{11}$, we can use Table \ref{tab:cont_table} to compute the remaining values. That is

\begin{equation}
    \pi_{10}=\pi_{1+}-\pi_{11}
\end{equation}

\begin{equation}
    \pi_{01}=\pi_{+1}-\pi_{11}
\end{equation}

\begin{equation}
    \pi_{00}=1+\pi_{11}-\pi_{1+}+\pi_{+1}
\end{equation}

To model continuous features, we assumed the features were normally distributed with a separate distribution for each outcome class. For $Y=0$, we modeled the feature as $\mathcal{N}(0,1)$ and for $Y=1$, we modeled the features as $\mathcal{N}(\mu,0)$ where $\mu$ was a parameter we controlled. By increasing $\mu$, the distributions would have less overlap, and hence, the value of a specific point would provide more information about the outcome.

For our study, we choose the same feature imbalances and odds ratio as found in \citeN{boulesteix_random_2012}, which correspond to $\pi_{+1}=\{0.5,0.4,0.25,0.1,0.05\}$ and $OR=\{3,1.5,1\}$, creating 15 binary features. Unlike \citeN{boulesteix_random_2012}, we chose to have a single feature with an odds ratio of 1 for each $\pi_{+1}$ rather than the 48 they use. We do so to keep the total number of features on the order of 10 rather than 100. We then created five continuous features with $\mu=\{0.75,0.50,0,0,0\}$, for a total of 20 features. We discuss our choice in the limitations.

\begin{figure}
 \includegraphics[width=1\linewidth]{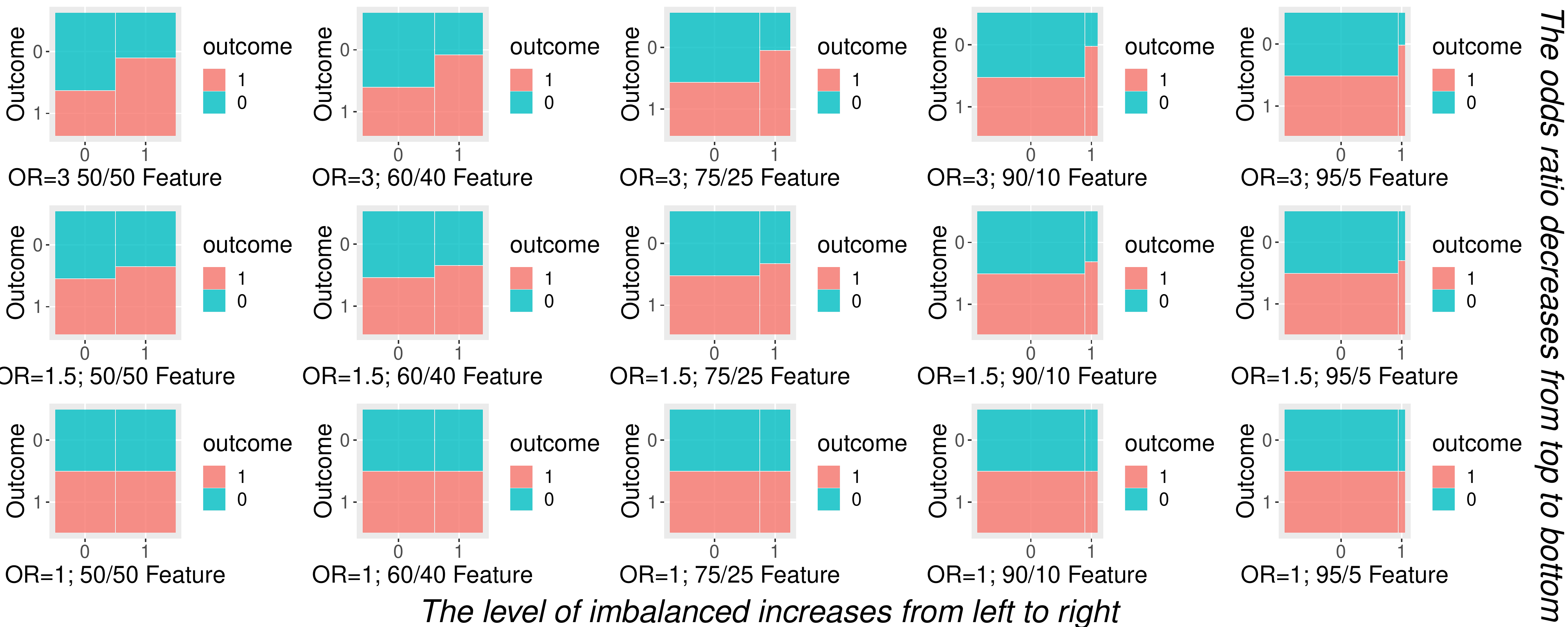}
 \caption{Distribution of binary features in the simulated $\pi_{1+}=0.5$, $N=1{,}000$ model. }
 \label{fig:binary_features}
\end{figure}

\begin{figure}
\centering
 \includegraphics[width=0.97\linewidth]{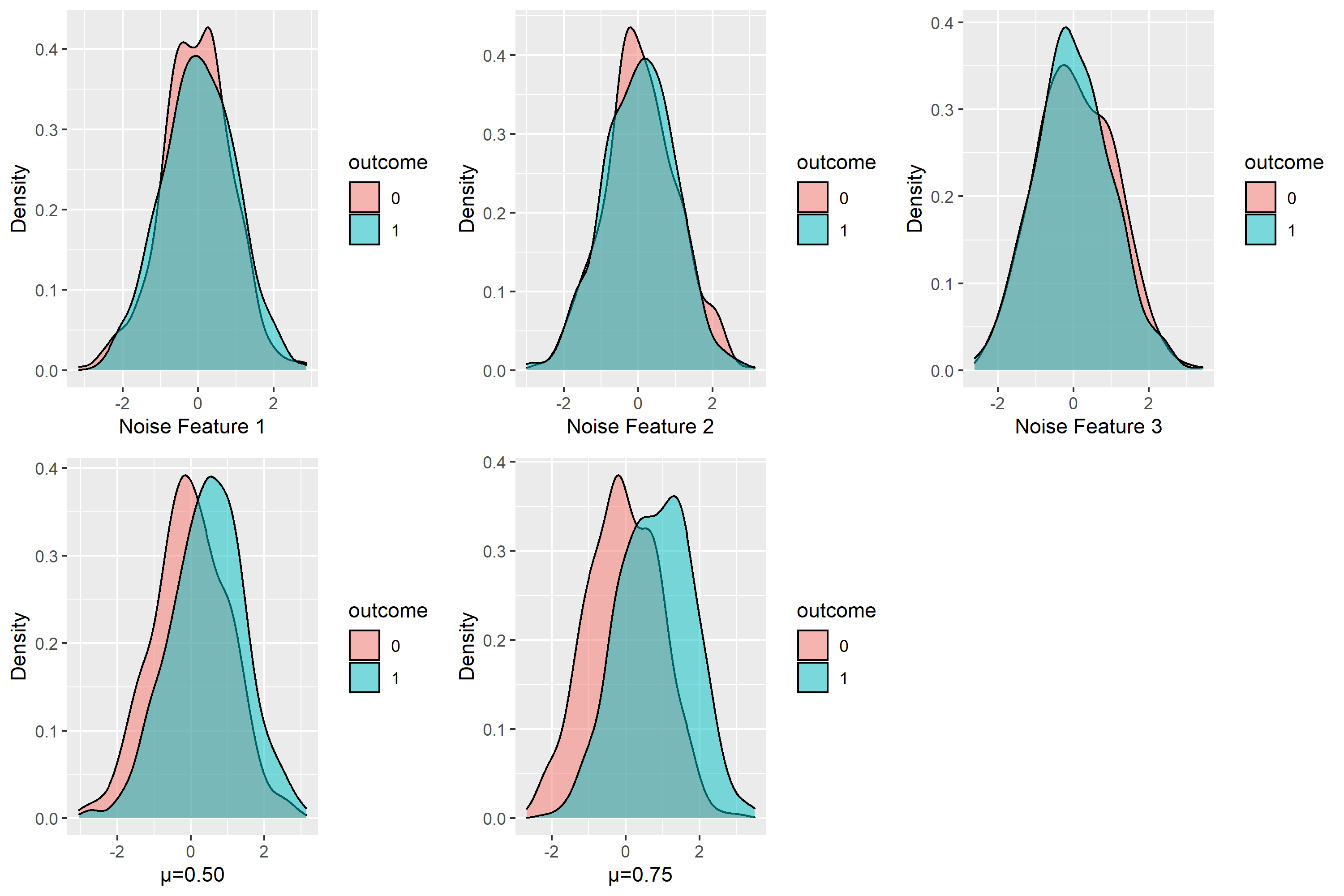}
 \caption{Distribution of continuous features in the simulated $\pi_{1+}=0.5$, $N=1,000$ model. }
 \label{fig:continuous_features}
\end{figure}

We then generated these features for five outcome imbalances, $\pi_{1+}=\{0.5,0.4,0.3,0.2,0.1\}$, and three sample sizes, $N=\{100,\; 1,000,\; 10,000\}$ for a total of 15 simulated data sets. A visual depiction of the binary features in the  $\pi_{1+}=0.5$ and $N=1,000$ case is shown in Fig.~\ref{fig:binary_features} and a visual depiction of the continuous features in that same case are shown in Fig.~\ref{fig:continuous_features}.

\subsection{Procedures}
\subsubsection{Forest Algorithms}
To analyze our data set using forest algorithms, we first randomly selected 70\% of the cases for the training set and kept the remaining 30\% of the data set for the testing set. Our prior work with random forest suggests that the size of the train/test split did not qualitatively affect the conclusions around variable importance and selected features \cite{young_identifying_2019}. We then used the \texttt{randomForest} function from the \texttt{randomForest} package \cite{liaw_classification_2002} to create random forest models and the \texttt{cforest} function from the \texttt{party} package \cite{hothorn_unbiased_2006,strobl_bias_2007,strobl_conditional_2008} to create conditional inference forests in R \cite{r_core_team_r:_2018}.

For both models, we set the number of trees to 500 as that is the default in the \texttt{cforest} algorithm and simulation studies of random forest have found that errors rates level off on the order of a few hundred trees \cite{svetnik_random_2003}. For the number of features per tree, we picked $\sqrt p$ where $p$ is the number of features, which is also aligned with the recommendations of \citeN{svetnik_random_2003}. We have called this $m$ previously to distinguish from probability in the logistic model but use $p$ here because it is the common symbol in the random forest literature.

For the random forest algorithm, we then computed the Gini importance. For the accuracy permutation importance and the conditional inference algorithm, we computed the AUC permutation importance and accuracy permutation importance. We repeated this procedure of splitting the data, running the model, and calculating the importances 30 times so that the resulting distribution of the importances would be approximately normal according to the central limit theorem \cite{heyde_central_2014}.

Next, we determined the rank of each feature based on its average value over the 30 runs, where the feature with the largest importance value would have rank 1. This type of approach is often used in screening studies to determine relevant features, which is what we are doing here \cite{hooker_please_2019}.

To evaluate bias in the \citeN{boulesteix_random_2012} paper, they approached bias as the difference from the expected value of zero in the null case and argued that bias when the features have odds ratios different 1 was not well defined. Because we are interested in selecting features, we can create a definition of bias based on the rank of the feature. If a forest algorithm is biased, we would expect to see that features with higher imbalance should have larger rank (i.e. be farther from 1) than features with identical odds ratios but smaller imbalances.

Using these ranks, we can also define bias in terms of the features detected by the algorithm. Assuming no bias, features with identical odds ratios should be detected at the same rate, regardless of their imbalance.

To determine if a feature was detected, we adopt the convention that detected means different from noise. We define detected as being ranked above the first noise feature, which has $OR=1$ or $\mu=0$. We picked this convention so that it is somewhat analogous to the definition statistically significant, which for explanatory models, is that the probability of obtaining a result at least as extreme as the result observed under the assumption of the null hypothesis is less than some threshold, typically 0.05.

\subsubsection{Regression Algorithms}
To use logistic regression in an explanatory manner, we did not use a train/test split as that approach is characteristic of a predictive approach and instead, used all of the data as is customary for explanatory modeling. To create a logistic regression model, we used the \texttt{glm} function that is part of base R with the option \texttt{family='binomial'} to use logistic instead of linear regression. Because log-F is based on data augmentation, we also used \texttt{glm} for that approach. Prior work suggests that a choice of $m=1$ performed better than a choice of $m=2$, and $m=1$ is a good starting choice when nothing is known about the size of the odds ratios \cite{rahman_performance_2017,greenland_penalization_2015}. Even though we ``know'' the true values of the odds ratios because we built them in, we want to approach the problem as if it were real data and we do not have any prior information about the features. We then used the default weights of $m/2$ for the log-F model.

To run the Firth penalization, we used the \texttt{brglm} function from the \texttt{brglm} package \cite{kosmidis_brglm_2020,kosmidis_jeffreys-prior_2020}. Per the function's documentation, the choice of \texttt{pl} is irrelevant for logistic regression so we left it at its default value.

For all three approaches, we used the \texttt{confint} function to compute the confidence intervals. For the Firth penalization, we picked \texttt{ci.method} to be \texttt{'mean'} as the \texttt{brglm} documentation suggests it is a less conservative approach. We then say that a feature was detected or statistically significant if zero is not in the confidence interval or in the case of odds ratios instead of the raw coefficients, 1 \cite{du_prel_confidence_2009}.

To get a sense of how the odds ratio varied based on the data, we also ran a bootstrapped simulation. That is, we randomly selected 80\% of the cases and ran the standard logistic regression, Firth penalization, and log-F penalization models on that data. We did this 10,000 times.

To create the lasso, ridge, and elastic net models, we again used a train/test split because these algorithms are designed for prediction rather than explanation. To align with the bootstrapping procedure, we used 80\% of the cases for the training data and 20\% for the testing data. Because lasso and ridge have a single tuning parameter, $\lambda$ that controls the amount of penalization, we used the \texttt{cv.glmnet} function to find the optimal value of $\lambda$. We then used the \texttt{glmnet} function to train the lasso and ridge models with their respective best value of $\lambda$ \cite{friedman_regularization_2010}. We again repeated this process 10,000 times.

Finally, we used the \texttt{train} function from the \texttt{caret} package to find the best values of $\alpha$ and $\lambda$ for elastic net and create the model \cite{kuhn_caret_2020}. We again did this 10,000 times with 80\% of the data used as training cases.

To analyze the bootstrapped results and generate confidence intervals for the values of the odds ratios, we used the percentile bootstraps \cite{efron_jackknife_1982}. Under this approach, all of the bootstrap estimates are sorted smallest to highest. For a given $\alpha$, the bootstrap confidence interval is interval lying between the $100 \times \frac{\alpha}{2}$ and $100 \times (1-\frac{\alpha}{2})$ percentiles. We chose $\alpha=0.05$ to form a 95\% bootstrapped confidence interval.

\subsection{Neutral Comparison Study Rationale}
Following the call of \citeN{boulesteix_plea_2013} for neutral comparison studies in the computational sciences, we address these three criteria and why we believe we have met their criteria.

\textit{A. The main focus of the article is the comparison itself. It implies that the primary goal of the article is not to introduce a new promising method.} As stated in the introduction, we are not introducing a method that we have developed and the focus of our paper is on comparing different methods rather than showing the usefulness of a certain method.

\textit{B. The authors should be reasonably neutral.} We have not developed any of the algorithms or techniques used in this study and hence, we have no stake in which method might perform best. We also have experience using predictive and explanatory methods and have used these techniques in our previous work.

\textit{C. The evaluation criteria, methods, and data sets should be chosen in a rational way,} Our methods and simulated data are based on a previously published simulated study, so we believe they are rational. We believe our evaluation criteria for detecting is rational because it is intuitive, objective, and based on prior approaches. We acknowledge that other approaches do exist and we address those in the discussion.

\section{Simulation Results}\label{sec:results}
\subsection{Forest Algorithm Results}

\begin{figure}
 \includegraphics[width=0.87\linewidth]{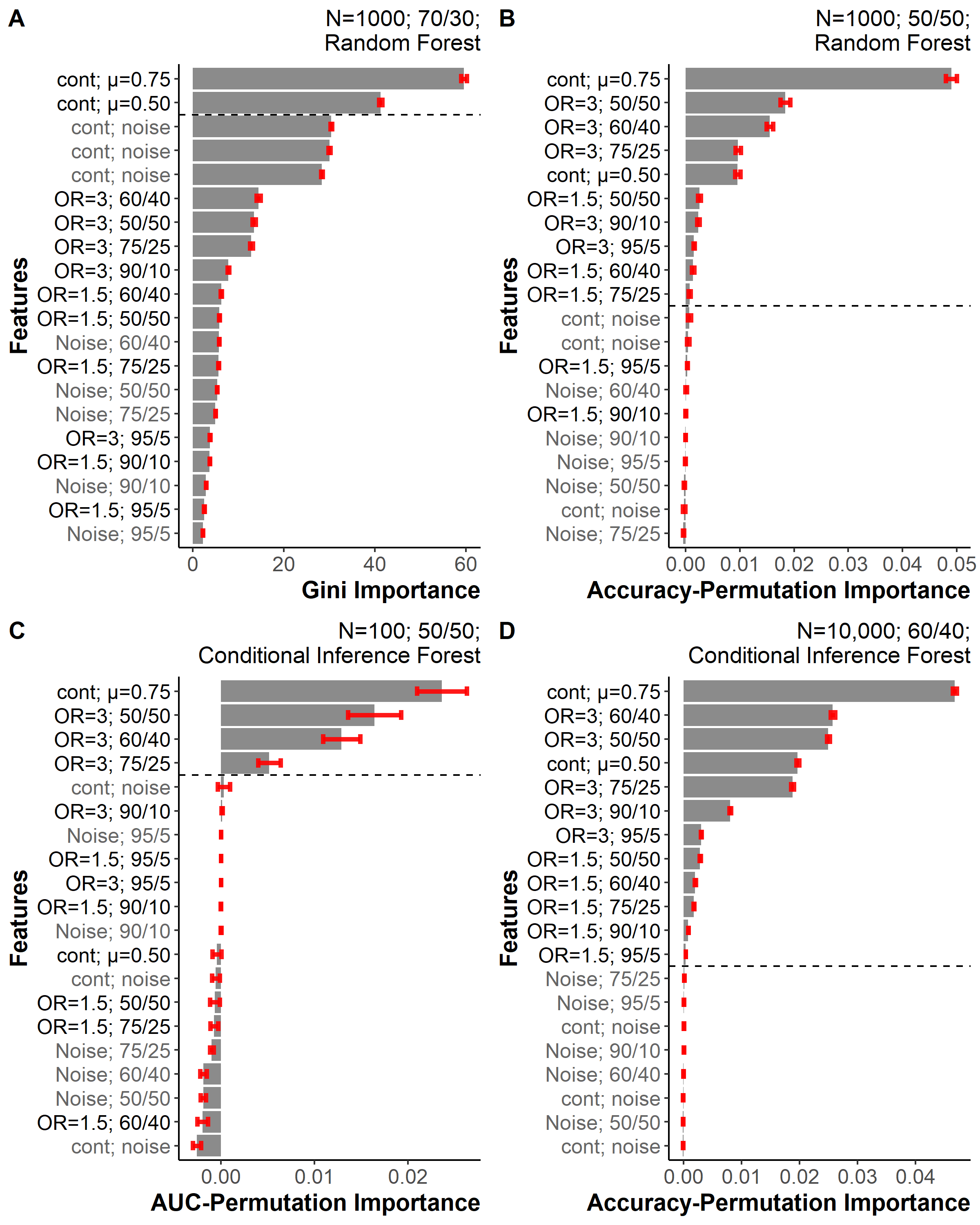}
 \caption{Importance values for a subset of the random forest models. Feature names shown in black were constructed to be informative while feature names in grey were constructed to be noise. Plot A shows the N=1000 70/30 outcome imbalance case with the standard random forest algorithm and Gini importance, plot B shows the N=1000 50/50 outcome imbalance case with the standard random forest algorithm and accuracy permutation importance, plot C shows the N=100, 50/50 outcome imbalance case with the conditional inference forest and AUC-permutation importance, and plot D shows the N=10,000 60/40 outcome imbalance case with conditional inference forest and accuracy-permutation importance. For all of the permutation importances, features with less imbalance tend to have larger importances than more imbalanced features for identical odds ratios.}
 \label{fig:RF_imp}
\end{figure}

\begin{figure}
    \centering
    \includegraphics[width=1\linewidth]{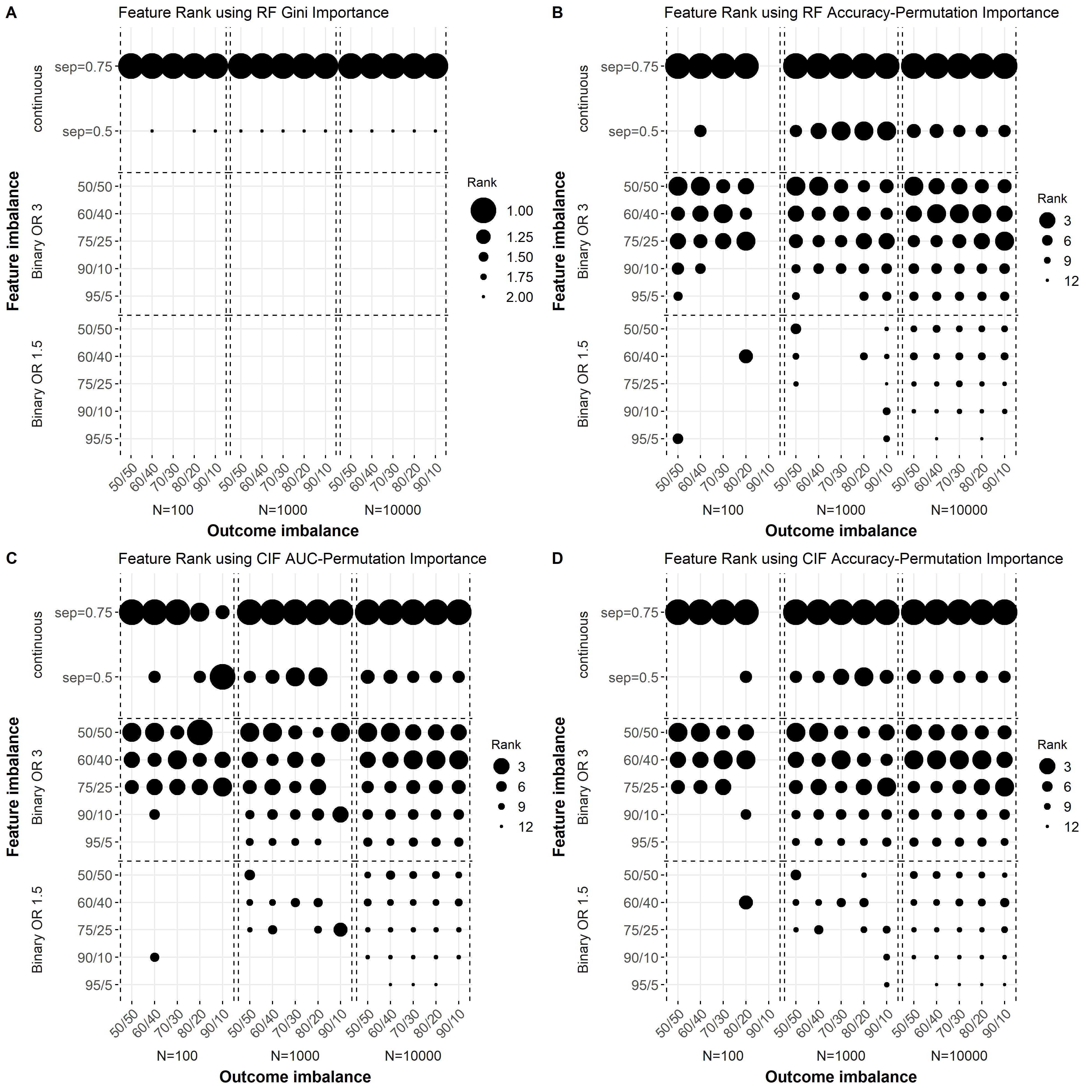}
    \caption{The ranks of the informative features for the four importance measures, grouped by the sample size and outcome imbalance. Any feature ranked below a noise feature was assigned a rank of 0 and hence results in no circle appearing the corresponding location. Here, a larger circle reflects a higher rank, meaning the feature was more predictive of the outcome. Overall, features with lower imbalance rank higher than features with higher imbalance for a given odds ratio and the result is not affected by the outcome imbalance or the specific permutation importance or forest algorithm used.}
    \label{fig:RF_bubble}
\end{figure}

When looking at a subset of the results in Fig.~\ref{fig:RF_imp}, we see similar results to \citeN{boulesteix_random_2012}. That is, for the Gini importance, represented by plot A, continuous features are ranked higher than binary features regardless of whether they are noise features or not. For the permutation importances, we see that more balanced features tend to have larger importances than less balanced features even when they have the same odds ratios. For example, when looking at Fig.~\ref{fig:RF_imp}D, we see that the features \textit{OR=3, 60/40} and \textit{OR=3, 50/50} have much higher importances than the \textit{OR=3, 90/10} and \textit{OR=3, 95/5} features. Similar trends are shown in plots B and C.

To compare across outcome imbalances, we aggregated all 3 sample sizes and 5 outcome imbalances into a single plot for each importance method. The results are shown in Fig.~\ref{fig:RF_bubble}.

Again, the Gini importance, Fig.~\ref{fig:RF_bubble}A, shows a preference toward continuous features and against binary features for all sample sizes and outcome imbalances. More specifically, there was not a single sample size or outcome imbalance in which any of the categorical features were detected.

For the permutation algorithms, the results are similar regardless of whether the accuracy-based or AUC-based permutation method is used. Regardless of outcome imbalance and sample size, more balanced features with smaller imbalances tend to rank higher than less balanced features with identical odds ratios. This result is reflected in the plots by the decreasing dot size from top to bottom in any of the rectangles formed by the dotted lines. In cases of high outcome imbalance, moderately imbalanced features might rank higher than the balanced feature (e.g. the \textit{Binary OR=3} features for the \textit{N=10,000 90/10} case in Fig.~\ref{fig:RF_bubble}D), but the most imbalanced features never rank higher than the balanced feature with the odds ratio. In fact, the \textit{OR=1.5 50/50} feature ranks higher than the \textit{OR=3, 95/5} feature for some of the models.

When looking at sections of columns of the plots in Fig.~\ref{fig:RF_bubble}, we notice that most informative features cannot be detected for $N=100$, regardless of which algorithm is used. In fact, only the less imbalanced \textit{OR=3} features and the more predictive continuous feature can be detected and even then, that depends on the level of outcome imbalance.

For the $N=1000$ case, most of the \textit{OR=3} features can be detected. However, only the less imbalanced \textit{OR=1.5} features are detected in most cases. Across the three permutation-based importances, there does not appear to be a consistent pattern for which \textit{OR=1.5} features are detected based on the outcome imbalance.

For the $N=10,000$ case, nearly all of the features can be detected, with the exception being the highly imbalanced \textit{OR=1.5 95/5} feature. Again, there is not a consistent pattern as to when this feature will not be detected based on the outcome imbalance. While the \textit{OR=1.5, 95/5} feature is never detected in the \textit{50/50} outcome imbalance, it is sometimes detected in the \textit{70/30} and \textit{90/10} outcome imbalance cases, making a pattern difficult to generalize based on the outcome imbalance.

\subsection{Logistic regression results}\label{sec:LR_results}
\begin{figure}
    \centering
    \includegraphics[width=1\linewidth]{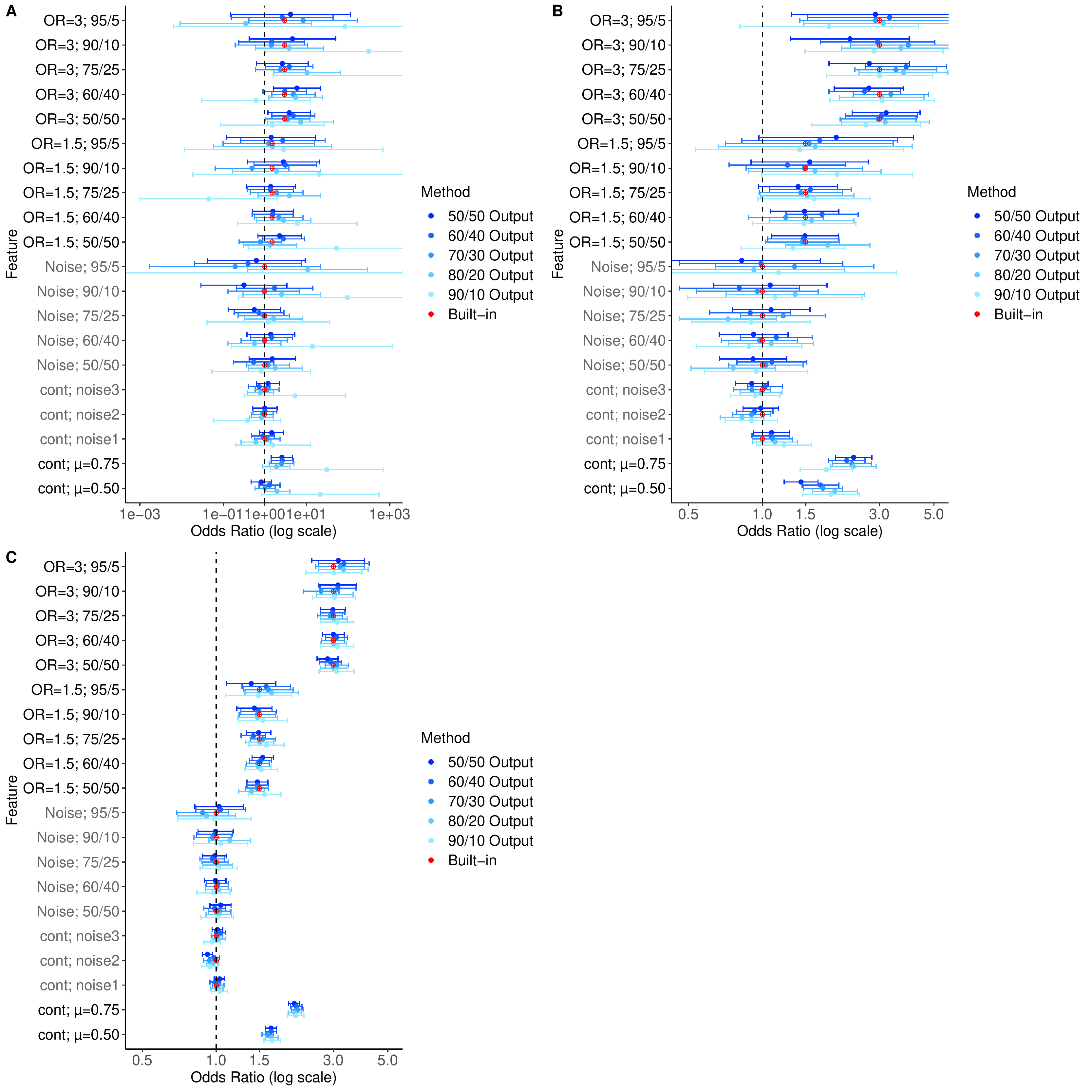}
    \caption{Values of the odds ratios and 95\% confidence intervals found by logistic regression models compared by outcome imbalance. Our built-in value is represented by the circled plus. Plot A is a sample size of $N=100$, plot B is a sample size of $N=1,000$ and plot C is a sample size of $N=10,000$. Confidence intervals without an end cap signify the error bar extends beyond the scale. Note the log scale on the horizontal axis and the different scale on (A) to show the error bars. In general, the confidence interval width increases with increasing feature and outcome imbalance and decreasing with increasing sample size.}
    \label{fig:lr_plot}
\end{figure}

In addition to detecting features, logistic regression provides an estimate of the odds ratio, which can give us an idea of how accurately algorithms are modeling the built-in odds ratios. Because the odds ratios and confidence intervals determine detection, we present those first. The odds ratio results are shown in Fig.~\ref{fig:lr_plot}.

From the $N=100$ case, plot A, we see that the 95\% confidence intervals for most features span at least an order of magnitude regardless of the feature imbalance or outcome imbalance. However, the width of the confidence interval tends to increase with both increasing feature imbalance and increasing outcome imbalance. For example, for the \textit{OR=3 90/10} and \textit{OR=3 95/5} features with a \textit{90/10} outcome imbalance, the confidence intervals are too wide to fit on a plot that spans 6 orders of magnitude. In some cases, the width of the confidence interval for a balanced feature with a highly imbalanced outcome can be comparable to a highly imbalanced feature with a balanced outcome such as \textit{OR=1.5 60/40} with a \textit{90/10} outcome imbalance and \textit{OR=3; 95/5} with a \textit{50/50} outcome balance.

Given the width of the confidence intervals, our built-in value of the odds ratio is always contained in the confidence intervals. However, when looking at the actual estimate of the odds ratio, we see varying degrees of accuracy. For some features, like \textit{OR=1.5; 50/50}, the \textit{80/20} outcome imbalance was the most accurate estimate while for \textit{OR=3; 60/40}, the \textit{60/40} outcome imbalance was the most accurate imbalance. In general, there was no specific trend where the discrepancy between the estimated value and the built-in value varied with increasing feature or outcome imbalance. In addition, there was no consistent trend where the estimated odds ratio over- or under-estimated the built-in value.

For the $N=1,000$ and $N=10,000$ cases, we notice that the confidence intervals have considerably shrunk and now span on the order of a single magnitude. This is true even for most imbalanced features with the most imbalanced outcome. Nevertheless, the width of the confidence intervals still tend to increase with both increasing feature imbalance and outcome imbalance.

As with the $N=100$ case, the built-in values are included in the confidence intervals and there is no consistent trend as to whether the estimated odds ratio over- or under-estimates the built-in value. We note that there is an exception to this for \textit{cont; noise2} and \textit{50/50} outcome imbalance on the $N=10,000$ plot where the noise feature is found to have an odds ratio less than 1.

\begin{figure}
    \centering
    \includegraphics[width=1\linewidth]{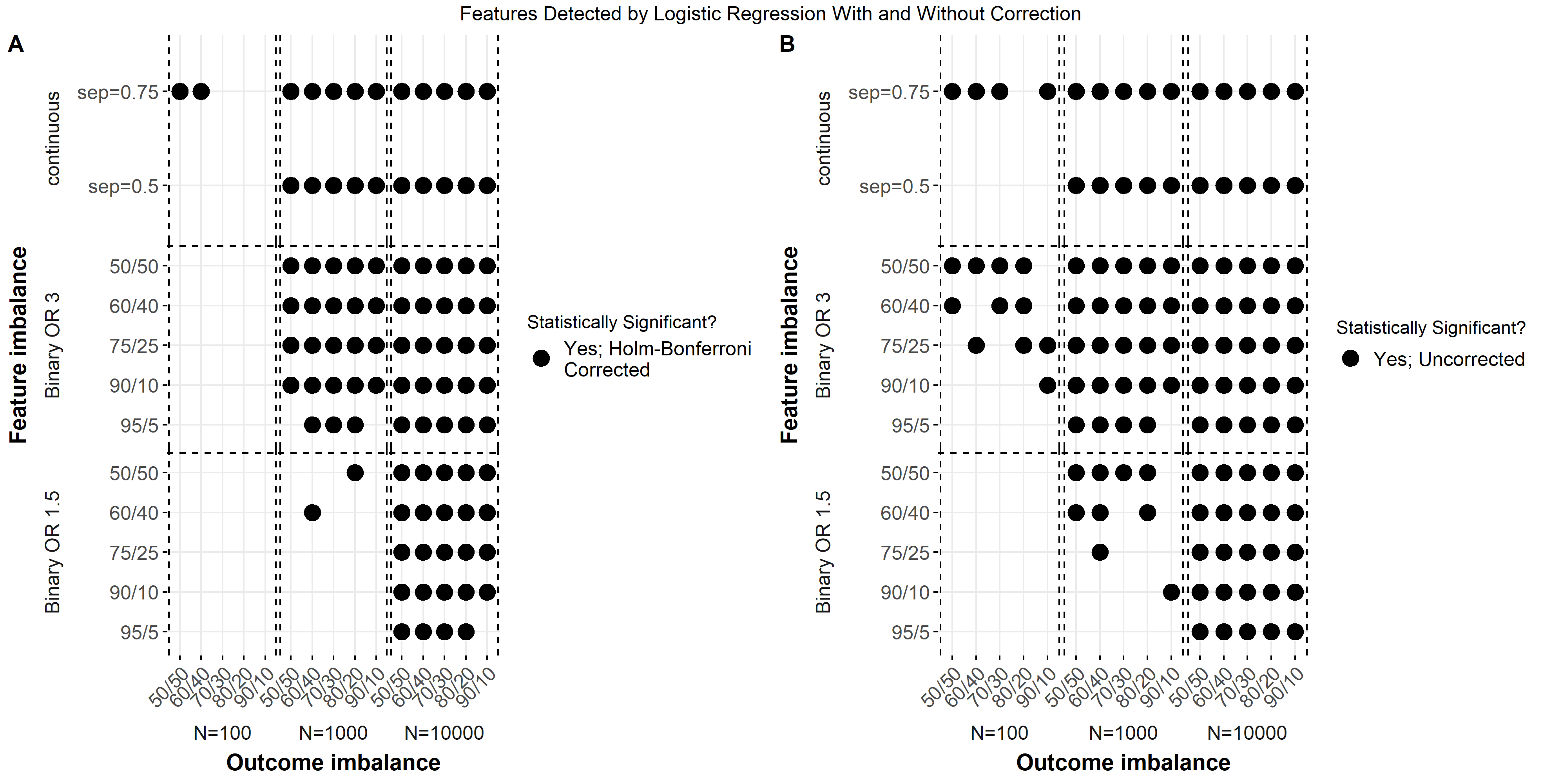}
    \caption{Analog of Fig.~\ref{fig:RF_bubble} but using logistic regression as the algorithm and statistical significance as the criteria for detection, $\alpha=0.05$. Plot A uses the Holm-Bonferroni correction to control for multiple tests while plot B uses the uncorrected p-values. Similar to  Fig.~\ref{fig:RF_bubble}, features with lower imbalance are more likely to be found as statistically significant than features with higher imbalance for a given odds ratio.}
    \label{fig:LR_features}
\end{figure}

Next, we can conduct an analysis similar to what we did with the forest algorithms and determine which features are detected by logistic regression. Here, because we are using logistic regression in an explanatory manner, we use the p-value to determine whether a feature is detected, with statistical significance meaning less than a chosen cutoff, $\alpha$. Because we are conducting multiple tests of statistical significance, we should control for false positives. Therefore, we present the results with and without a Holm-Bonferroni correction \cite{holm_simple_1979}, which is less conservative than the traditional Bonferroni correction. The correction is applied within each data set because for a study with real data, we would only have one data set. The results are shown in Fig.~\ref{fig:LR_features}

For $N=100$, when we apply the Holm-Bonferroni correction, the continuous feature with the largest $\mu$ is the only one to be detected and even then, only for minor outcome imbalances. If instead we do not apply any corrections, logistic regression is able to detect a few of the \textit{OR=3} features however these tend to be the ones with lower imbalances. That is, even with a generous definition of statistical significance, logistic regression is unable to detect features with moderate odds ratios or features with large odds ratios but higher imbalances.

For $N=1000$, logistic regression is able to detect both continuous features and most of the \textit{OR=3} features regardless of whether we applied a correction to the p-values or not. Unlike the $N=100$ case, we are able to detect some of the \textit{OR=1.5} features though only features with lower imbalances and this depends on whether we apply a correction or not. When we apply the correction, we were only able to detect two of the \textit{OR=1.5} features across any of the five outcome imbalances, while if we did not apply the correction, we could detect ten.

Finally, for $N=10,000$, we were able to detect all of the informative features, regardless of whether we applied a correction or not. However, one of the continuous noise features was marked as statistically significant in the \textit{50/50} outcome imbalance and the \textit{70/30} outcome imbalance cases. One of these disappeared when we applied the p-value correction while one did not, suggesting that with enough data, random variations in the data might appear as signals.

\subsection{Penalized regression results}
Given the result from Sec.~\ref{sec:LR_results} that most features are detected for $N=10,000$ even without correction, we chose to focus on the $N=100$ and $N=1000$ cases as areas where penalized regression might offer a benefit. To get a representative picture of how penalized regression might help, we then applied the algorithms to the \textit{50/50}, \textit{70/30}, \textit{90/10} imbalanced outcome data sets, representing no imbalance, medium imbalance, and high imbalance.

\subsubsection{Confidence interval approach}

\begin{figure}[hbt!]
    \centering
    \includegraphics[width=1\linewidth]{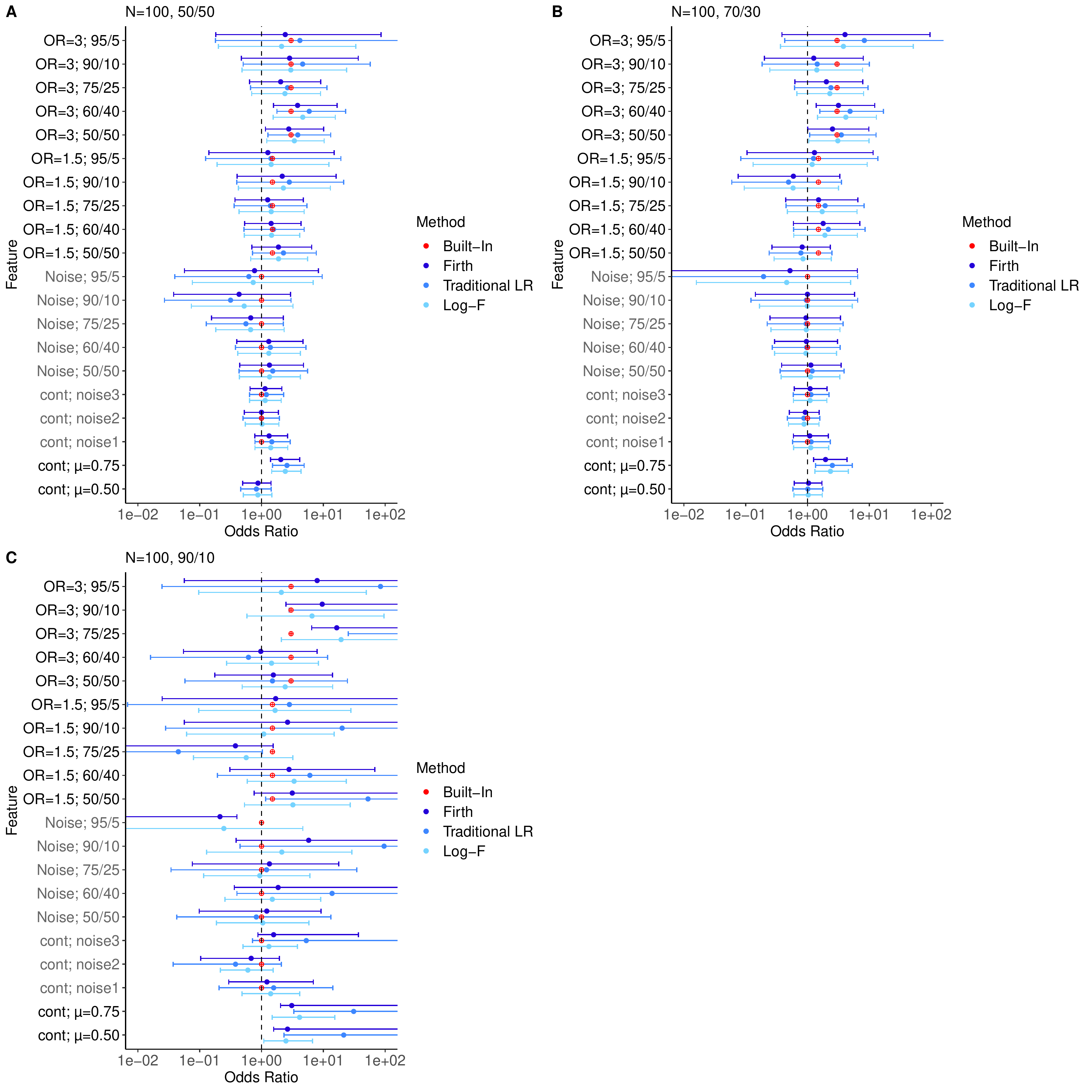}
    \caption{95\% confidence intervals for Firth penalized, traditional, and Log-F penalized logistic regression for the $N=100$ data sets. Plot A shows the \textit{50/50} outcome imbalance, plot B shows the \textit{70/30} outcome imbalance, and plot C shows the \textit{90/10} outcome imbalance. Confidence intervals that span beyond the scale are removed from the plot. For higher outcome imbalance, Firth and Log-F penalizations can considerably shrink the confidence intervals.}
    \label{fig:ci_LR_n100}
\end{figure}

\begin{figure}[hbt!]
    \centering
    \includegraphics[width=1\linewidth]{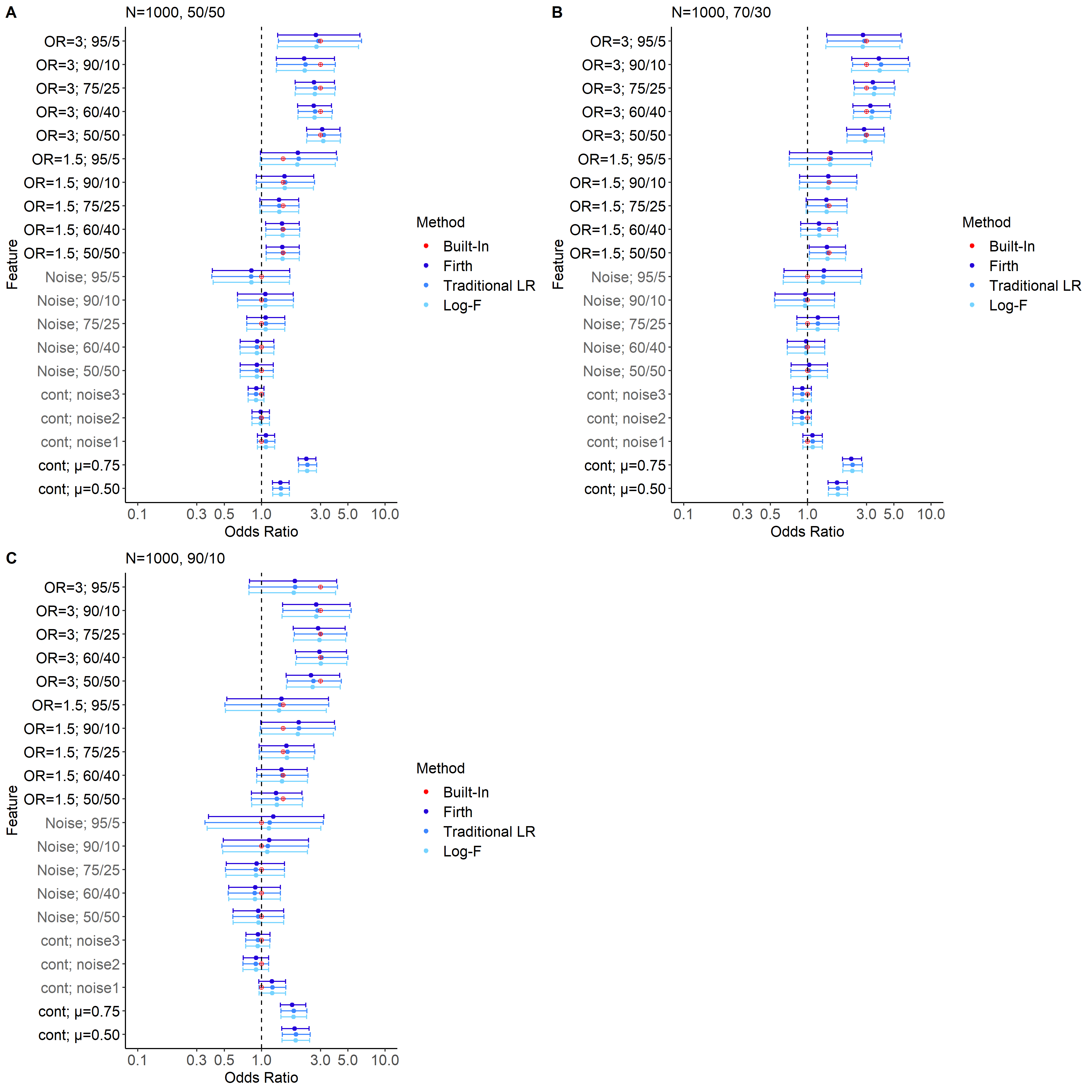}
    \caption{95\% confidence intervals for Firth penalized, traditional, and Log-F penalized logistic regression for the $N=1000$ data sets. Plot A shows the \textit{50/50} outcome imbalance, plot B shows the \textit{70/30} outcome imbalance, and plot C shows the \textit{90/10} outcome imbalance. Confidence intervals that span beyond the scale are removed from the plot. For higher outcome imbalance, Firth and Log-F penalizations can shrink the confidence intervals.}
    \label{fig:ci_LR_n1000}
\end{figure}

Because Firth and Log-F penalized regression are designed for explanatory approaches, we can use them to generate confidence intervals. The results for the $N=100$ data sets are shown in Fig.~\ref{fig:ci_LR_n100} and the results for the $N=1000$ data sets are shown in Fig.~\ref{fig:ci_LR_n1000}. Here, we only present the uncorrected 95\% confidence intervals because if we do not find a benefit on the uncorrected confidence intervals, we would not find one on the corrected versions.

For the $N=100$ case, we notice that the Firth and Log-F penalizations tend to have smaller confidence intervals and in many cases, are closer to the built-in odds ratio than traditional logistic regression is.

For the \textit{50/50} case, all three algorithms produce similar confidence intervals for more balances features such as \textit{OR=3; 50/50}. For the highly skewed features such as \textit{OR=3; 95/5}, Firth and Log-F penalizations do shrink the confidence interval compared to the traditional method with Log-F appearing to offer a greater benefit. However, none of the shrinking makes a difference as to whether the feature would be statistically significant or not compared to traditional logistic regression.

When we instead look at the moderately imbalanced \textit{70/30} case, we see similar results. That is, the Firth and Log-F penalizations appear to provide a greater benefit in terms of shrinking the confidence interval for features with greater imbalance, though again, the benefit is not enough to change whether a feature would be detected.

For the highly imbalanced \textit{90/10} case, both penalizations reduce the confidence intervals regardless of the feature's imbalance. The benefits are most clear however for the most imbalanced features. For example, for \textit{OR=3; 90/10}, Log-F penalization reduces the width of the confidence interval by nearly 3 orders of magnitude compared to the traditional logistic regression. As in the \textit{50/50} and \textit{70/30} cases, the penalizations do not affect whether a feature would be statistically significant, but the penalizations still do produce more accurate estimates of the built-in odds ratios than traditional logistic regression does.

Looking at the $N=1000$ results in Fig.~\ref{fig:ci_LR_n1000}, we notice that the confidence intervals of the penalized regression methods are similar in length to those of traditional logistic regression. This result is true regardless of the feature imbalance or the outcome imbalance.

When it comes to estimating our built-in odds ratio, the penalized methods do not offer much of an improvement over traditional logistic regression. Indeed, for lower imbalanced features, all three methods tend to provide similar estimates, while for higher imbalanced features, there is no clear trend as to which method will provide an estimate closest to that of the built-in value.

\subsubsection{Bootstrap approach}

\begin{figure}[hbt!]
    \centering
    \includegraphics[width=\linewidth]{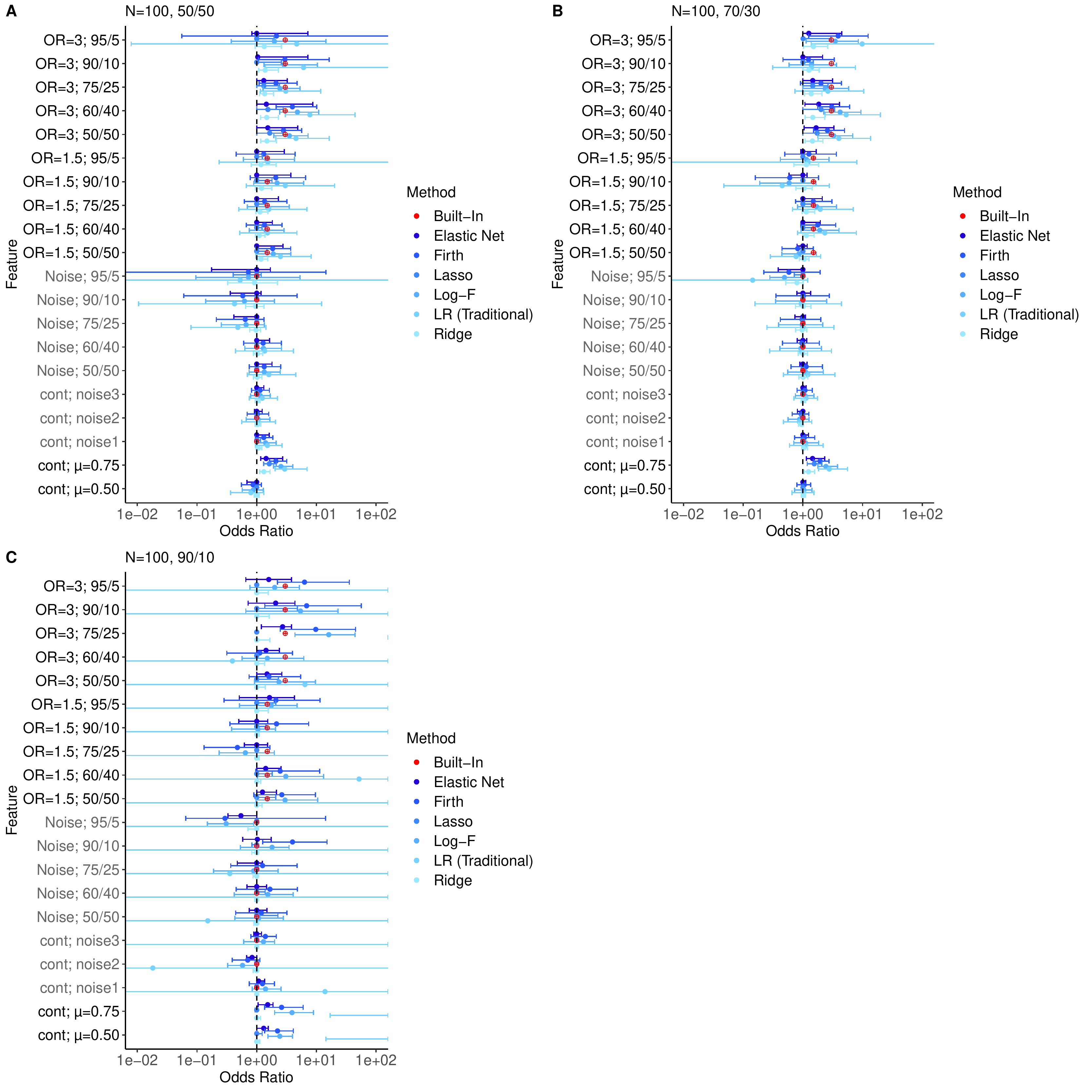}
    \caption{95\% percentile bootstraps of the odds ratio for elastic net, Firth, lasso, Log-F, no, and ridge penalizations on the $N=100$ data. Dots represent the median value. Plot A shows the \textit{50/50} outcome imbalance, plot B shows the \textit{70/30} outcome imbalance, and plot C shows the \textit{90/10} outcome imbalance. Confidence intervals without an end cap signify the error bar extends beyond the scale. In general, traditional logistic regression has the widest confidence intervals}
    \label{fig:bootstrap_LR_n100}
\end{figure}

\begin{figure}[hbt!]
    \centering
    \includegraphics[width=1\linewidth]{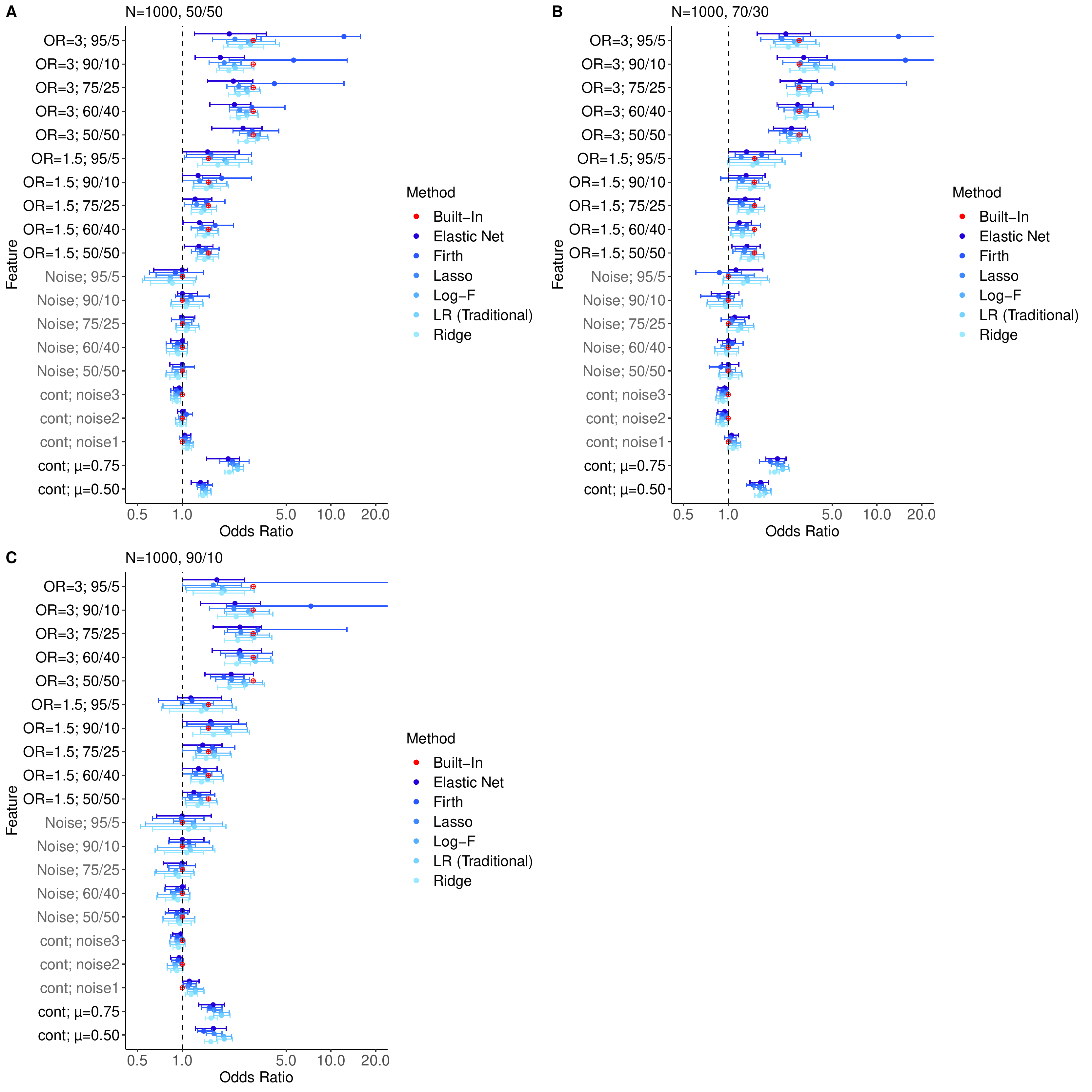}
    \caption{95\% percentile bootstraps of the odds ratio for elastic net, Firth, lasso, Log-F, no, and ridge penalizations on the $N=1,000$ data. Dots represent the median value. Plot A shows the \textit{50/50} outcome imbalance, plot B shows the \textit{70/30} outcome imbalance, and plot C shows the \textit{90/10} outcome imbalance. Confidence intervals without an end cap signify the error bar extends beyond the scale. Most methods perform similar with the exception of Firth penalization at high feature imbalance.}
    \label{fig:bootstrap_LR_n1000}
\end{figure}

In addition to only considering whether the algorithm detects a feature, we can also get a sense of what range the estimated odds ratio will fall in using the five different penalization approaches. The results from the $N=100$ data sets are shown in Fig.~\ref{fig:bootstrap_LR_n100} and the results from the $N=1000$ data sets are shown in Fig.~\ref{fig:bootstrap_LR_n1000}.

From the $N=100$ plot, we see that spread of the estimated values varies between the different methods. For higher feature imbalances, traditional logistic regression and Firth penalized regression often have the widest distributions. Because lasso shrinks the coefficients to zero (or equivalently, odds ratios to 1) and ridge reduces the variance of the estimate, these two methods often have the most compact distributions.

Likewise, in terms of the median estimate of the odds ratio, we see variation between the methods. Because lasso shrinks estimates and ridge scales estimates, these two underestimate the built-in odds ratio. We also find this behavior with elastic net, which is a middle group between the two. However, elastic net often includes the built-in odds ratio within its interval even when lasso and ridge do not. This result is especially true for higher feature and outcome imbalances.

Log-F penalization on the other hand often takes a middle ground on both estimates and distribution width. Regardless of the feature or outcome imbalance, Log-F does not consistently over- or under-estimate the built-in odds ratio and does not have the widest distribution of the estimates.

From the $N=1000$ results shown in Fig.~\ref{fig:bootstrap_LR_n1000}, we see that the six methods tend to produce similar results for more balanced features, even at higher outcome imbalances. The exception is the Firth penalization for higher imbalance features (e.g. \textit{OR=3; 90/10}). For these higher imbalance features, the Firth penalization estimates can be nearly an order of magnitude larger than the estimates produced by other methods.

As in the $N=100$ case, we find that lasso and ridge tend to underestimate the built-in odds ratio for the $N=1000$ case. Unlike the $N=100$ case, however, the built-in value is included in the bootstrapped confidence interval.

In most cases, elastic net, Log-F, and logistic regression tend to have similar distribution widths and have the built-in odds ratios within their intervals. While elastic net under-predicts the built-in value, Log-F penalized and traditional logistic regression do not show a consistent pattern as to whether they over- or under-predict the built-in value.

\section{Application to Real Data}\label{sec:realdata}
In this section, we apply the results of our simulation study to a graduate admissions data set.

Our data set comes from the application records of over 5,000 applicants to the physics graduate program at 6 Big Ten or Midwestern universities over a two-year period. The data includes the applicant's GRE scores, undergraduate GPA, undergraduate university, demographics such as binary gender, race, domestic status, whether the applicant made the shortlist, and whether the applicant was admitted to the program. Details about these features can be found in \citeN{posselt_metrics_2019}.

We can then treat each of the six physics graduate programs as a separate case study, which is an approach we have used in our previous work \cite{young_using_2020}. Doing so allows us to vary the sample size and the outcome imbalance. For the six graduate programs in the data set, the smallest program had $N=140$ applicants over the two year period while the largest had $N=1228$. When considering whether the applicant made the shortlist or was admitted, the outcome imbalance ranged from $53/47$ to $83/17$, which means that the sample size and outcome imbalances are on the same scale as the data we used in our simulation study.

We then selected a subset of the programs that represent a small and medium data set with a more balanced and less balanced outcome. Specifically, we modelled school 1's admission (N=140, 59/41), school 2's shortlist (N=431, 78/22), and school 3's shortlist and admission (N=1228, 60/40 and 78/22) respectively. In the initial paper using this data, \citeN{posselt_metrics_2019} analyzed shortlist and admissions separately and hence, we do so here.

\subsection{Methods}
\begin{table}[]
\centering
\caption{Feature and outcome imbalances for the binary features from actual graduate school admission data. Numbers in the table refer to the percent of cases in each category of the binary feature.}
\label{tab:igen_imbalance}
\begin{tabular}{ccccc}
\hline
Feature     & \multicolumn{4}{c}{School}    \\
 &
  \begin{tabular}[c]{@{}l@{}}School 1 \\ Admit\end{tabular} &
  \begin{tabular}[c]{@{}l@{}}School 2 \\ Admit\end{tabular} &
  \begin{tabular}[c]{@{}l@{}}School 3\\ Shortlist\end{tabular} &
  \begin{tabular}[c]{@{}l@{}}School 3\\ Admit\end{tabular} \\ \hline
Outcome     & 59/41 & 83/17 & 59/41 & 76/24 \\
Gender      & 79/21 & 81/19 & 85/15 & 85/15 \\
Domestic    & NA    & 71/29 & 50/50 & 50/50 \\
Year        & 57/43 & 55/45 & 51/49 & 51/49 \\
Race=Asian  & 87/13 & 64/36 & 52/48 & 52/48 \\
Race=Black  & 96/4  & 99/1  & 99/1  & 99/1  \\
Race=Latinx & 81/19 & 91/9  & 99/1  & 99/1  \\
Race=Multi  & 96/4  & 97/3  & 93/7  & 93/7  \\
BinaryNoise1& 60/40 & 60/40 & 60/40 & 60/40 \\
BinaryNoise2& 75/25 & 75/25 & 75/25 & 75/25 \\
BinaryNoise3& 90/10 & 90/10 & 90/10 & 90/10 \\
BinaryNoise4& 95/5  & 95/5  & 95/5  & 95/5  \\ \hline
N           & 140   & 431   & 1228  & 1228  \\ \hline
            &       &       &       &
\end{tabular}
\end{table}

To analyze the real data, we used five approaches. First, we use logistic regression and random forest with the Gini importance as they are the ``default'' methods. Based on the results of the simulation study, we then choose to use Log-F, as it performed either better or no worse than Firth, elastic net, as it performed better than lasso or ridge and retains the benefits of both, and conditional inference forest with the AUC importance, as all of the permutation based importance measures performed similarly.

To mimic the simulation study and know which features were certainly noise, we added four binary noise features (imbalances of 60/40, 75/25, 90/10, and 95/5, which we refer to as BinaryNoise1, BinaryNoise2, BinaryNoise3, BinaryNoise4) and three continuous noise features. The binary features and their imbalances for the four data sets are shown in Table \ref{tab:igen_imbalance}.

To run the models, we used the same R packages as in the simulation study. However, for real data, we should be interested in how well the model fits and hence, need to include some measure of that. For the logistic regression based methods, we used the standard McFadden pseudo-$R^2$ implemented in the \texttt{DescTools} package via the \texttt{PseudoR2} function \cite{signorell_desctools_2020}, where a good value is between 0.2 and 0.4 \cite{mcfadden_quantitative_1977}. While other choices of pseudo-$R^2$ exist, \citeN{menard_coefficients_2000} suggests that there is little reason to prefer one over another, but McFadden's might be preferable because it is intuitive.

To connect the forest methods with the logistic regression methods, we also computed the AUC for each model, which follows the recommendation of \citeN{aiken_framework_2021}. To do so, we used the \texttt{AUC} function from the \texttt{ModelMetrics} package \cite{hunt_modelmetrics_2020}. We interpreted an AUC of at least 0.7 as a good model \cite{araujo_validation_2005}.

For the predictive methods, elastic net, random forest, and conditional inference forest, we used the same procedure as in the simulation study except now used a 80/20 train/test split for all methods and calculated the AUC on both the training and testing data sets.

\subsection{Results}
\begin{table}[]
\centering
\caption{McFadden Pseudo $R^2$ values for the explanatory models}
\label{tab:mcfadden_table}
\begin{tabular}{ccccc}
\hline
                    & School 1 & School 2 & School 3 shortlist & School 3 admit \\ \hline
Logistic Regression & 0.256    & 0.215    & 0.199              & 0.203          \\
Log-F               & 0.252    & 0.215    & 0.199              & 0.203          \\ \hline
\end{tabular}
\end{table}

\begin{table}[]
\centering
\caption{AUC values for the various models on the four data sets}
\label{tab:auc_table}
\begin{tabular}{lllll}
\hline
                                     & School 1 & School 2 & School 3 shortlist & School 3 admit \\ \hline
Logistic Regression                  & 0.826    & 0.806    & 0.790              & 0.799          \\
Log-F                                & 0.825    & 0.805    & 0.790              & 0.799          \\
Elastic (Train)                      & 0.830    & 0.807    & 0.785              & 0.799          \\
Elastic (Test)                       & 0.690    & 0.734    & 0.771              & 0.779          \\
Random Forest (Train)                & 0.547    & 0.529    & 0.688              & 0.613          \\
Random Forest (Test)                 & 0.564    & 0.521    & 0.686              & 0.616          \\
Conditional Inference Forest (Train) & 0.749    & 0.517    & 0.793              & 0.674          \\
Conditional Inference Forest (Test)  & 0.594    & 0.500    & 0.681              & 0.597          \\ \hline
\end{tabular}
\end{table}

\begin{figure}
    \centering
    \includegraphics[width=1\linewidth]{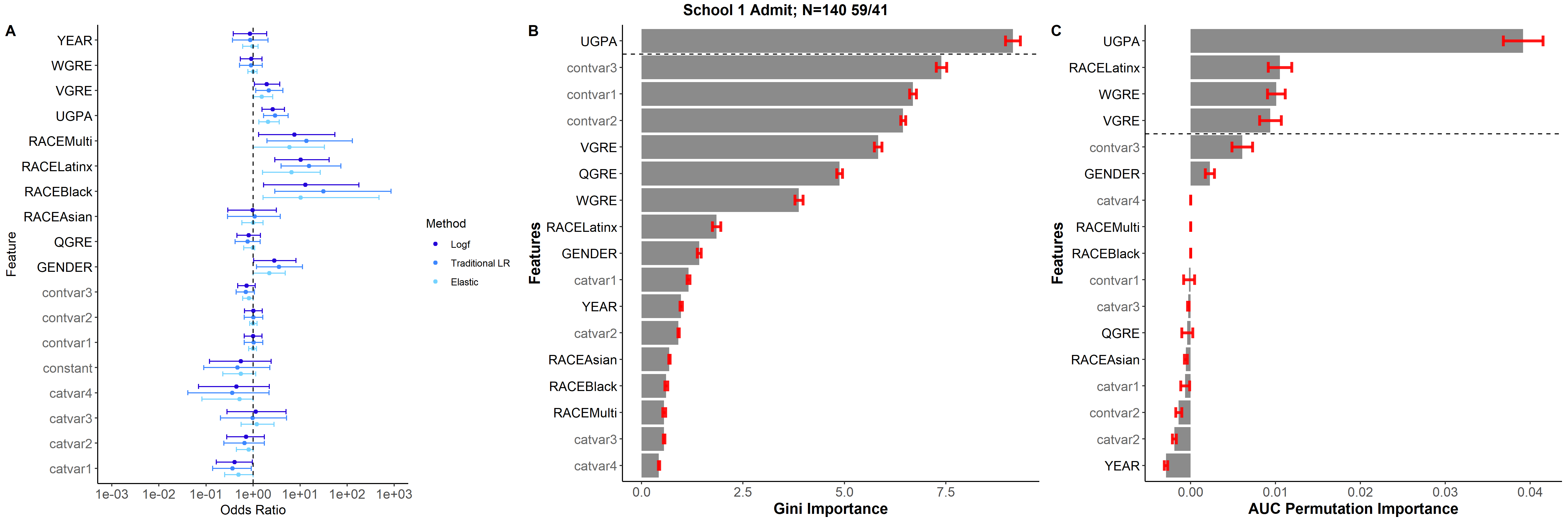}
    \caption{Comparison of the odds ratio (A), Gini importance (B), and AUC-permutation importance (C) for the features in school 1. Notice that RaceLatinx has a similar odds ratio as RaceBlack and RaceMulti according to (A) but only RaceLatinx is detectable in (C). RaceLatinx is less imbalanced than RaceBlack and RaceMulti.}
    \label{fig:school1}
\end{figure}

First, we present the metrics used to assess our model, which are shown in Table \ref{tab:mcfadden_table} and Table \ref{tab:auc_table}. We notice that except for school 3 shortlist, all of the pseudo $R^2$ are within the accepted range ($\geq 0.2$).

When looking at the AUC values, we notice that the regression models outperform the forest models and in most cases, the forest models do not produce an AUC in the acceptable range. As our goal is not to make the best model but rather to extract features, we did not do any parameter tuning for the forests. We discuss more about these metrics in the discussion.

Because the conclusions from the four data sets are similar, we share only the results of school 1 and provide plots for the other data sets in the appendix for completeness. The results of algorithms applied to the school 1 data set are shown in Fig.~\ref{fig:school1}.

When looking at plot A, we notice that Log-F noticeably shrinks the confidence interval for highly skewed features like \textit{RaceBlack}. In exchange though, the estimate of the odds ratio is shrunk closer to $OR=1$ for nearly all the features.

Even though elastic net is showing the percentile bootstrapped confidence interval instead of the statistical confidence interval, the results tend to be aligned with the other methods. That is, the median value is on the same order of magnitude of the other estimates and the end points of the confidence interval are also on the same order of the magnitude as the other estimates.

When comparing the different methods, we see that none of the three algorithms would have led to different conclusions about which features are statistically significant or not. From plot A, the statistically significant features would be \textit{VGRE}, \textit{UGPA}, \textit{RaceMulti}, \textit{RaceLatinx}, \textit{RaceBlack}, and \textit{BinaryNoise1}. In the case of \textit{BinaryNoise1}, which is supposed to be a noise feature, we note that due to the random nature of generating the feature, the odds ratio was smaller than 1 and hence, the algorithms appear to have detected that small difference.

When we move to plot B, we note that the continuous features are all ranked above the binary features as expected. As a result, all features except for one rank lower than the first noise feature.

Finally, when we move to plot C, we notice that only four features are detected, which is smaller than the regression approaches. Because prediction and explanation have different goals, we would not expect them to identify the same features. Yet, multiple approaches identifying the same features suggest that these features are in fact, distinct from noise.

One interesting point to note is that we see some ranking issues based on imbalance. For example, using a 2x2 contingency table to calculate the theoretical odds ratios, \textit{RaceMulti} should have an odds ratio of 2.96 while \textit{RaceLatinx} should have an odds ratio of 2.25. However, because \textit{RaceLatinx} has an imbalance of 80/20 while \textit{RaceMulti} has an imbalance of 96/4, \textit{RaceLatinx} is detected by the AUC-permutation importance while \textit{RaceMulti} is not.

\section{Discussion}\label{sec:discussion}
Here we address our research questions and consider how our choices and approaches might have impacted the conclusions we can draw from this study. We include a summary of the advantages and disadvantages of each algorithm based on our study and prior work in Table \ref{tab:algorithm_pros_cons}.

\begin{landscape}
\begin{table}
\caption{Summary of advantages and disadvantages for each algorithm used in this study}
\label{tab:algorithm_pros_cons}
\begin{tabular}{lll}
\hline
\multicolumn{1}{c}{Method} &
  \multicolumn{1}{c}{Advantages of algorithm} &
  \multicolumn{1}{c}{Disadvantages of algorithm} \\ \hline
RF + Gini &
  \begin{tabular}[c]{@{}l@{}}-Default choice for many random forest\\ implementations\end{tabular} &
  \begin{tabular}[c]{@{}l@{}}-Biased in favor of continuous features, regardless\\ of whether they are informative of the outcome or\\ not\end{tabular} \\
\begin{tabular}[c]{@{}l@{}}RF + accuracy \\ permutation importance,\\ CIF + accuracy \\ permutation importance,\\ CIF+ AUC permutation \\ importance\end{tabular} &
  \begin{tabular}[c]{@{}l@{}}-Can be used with continuous and categorical features\\ -Categorical features do not need to be binarized\\ -Comparable performance to logistic regression for \\ feature selection without needing to check any \\ assumptions\end{tabular} &
  \begin{tabular}[c]{@{}l@{}}-Ability to detect features decreases with increasing \\ feature imbalance and outcome imbalance\\ -Questionable performance for small N\end{tabular} \\
Logistic Regression &
  \begin{tabular}[c]{@{}l@{}}-standard algorithm for classification, implemented in\\ most software\\ -odds ratios have a ``real-world'' interpretation\end{tabular} &
  \begin{tabular}[c]{@{}l@{}}-Width of confidence interval increases for\\ increased outcome and feature imbalance and can\\ become infinite in some cases\end{tabular} \\
Firth penalization &
  \begin{tabular}[c]{@{}l@{}}-Able to shrink confidence intervals for imbalanced \\ features in small N situations\end{tabular} &
  \begin{tabular}[c]{@{}l@{}}-Not widely implemented in software\\ -Advantages compared to logistic regression \\ disappear for larger N\end{tabular} \\
Log-F penalizations &
  \begin{tabular}[c]{@{}l@{}}-Able to shrink confidence intervals for imbalanced \\ features in small N situations\\ -Based on data augmentation so no special software\\ needed\\ -Coefficient estimates are similar to those of \\ traditional logistic regression\end{tabular} &
  \begin{tabular}[c]{@{}l@{}}-Advantages compared to logistic regression \\ disappear for larger N\end{tabular} \\
Lasso &
  \begin{tabular}[c]{@{}l@{}}-Shrinks some coefficients to zero which can be \\ useful for feature selection\end{tabular} &
  \begin{tabular}[c]{@{}l@{}}-Less able to detect less informative features from\\ noise compared to other algorithms in the study\end{tabular} \\
Ridge &
  \begin{tabular}[c]{@{}l@{}}-Effective at shrinking the width of the distribution\\ of estimated odds ratios\end{tabular} &
  \begin{tabular}[c]{@{}l@{}}-All coefficients are scaled by the same amount\\ and are underestimated\end{tabular} \\
Elastic Net &
  \begin{tabular}[c]{@{}l@{}}-Combines the benefits of lasso and ridge\\ penalizations, often performing better than either \\ approach individually\end{tabular} &
  \begin{tabular}[c]{@{}l@{}}-Requires hyperparameter tuning to determine the\\ ideal amount of mixing between lasso and ridge\end{tabular} \\ \hline
\end{tabular}
\end{table}
\end{landscape}

\subsection{Research Questions}

\textit{How might known random forest feature selection biases change when the outcome is imbalanced as is often the case in EDM studies, and does the AUC-permutation importance affect those biases?} When we vary the outcome imbalance as well as the feature imbalance, we still observe the same general trend as seen in \citeN{boulesteix_random_2012}. That is, features with higher imbalance are less likely to be detected compared to features with lower imbalances but the same odds ratio. In fact, the bias might become worse for high outcome imbalances because it is harder to train a ``good'' model when most of the cases have the same outcome.

In opposition to the claims of \citeN{janitza_auc-based_2013}, we do not find the AUC permutation importance to outperform the accuracy permutation importance. In fact, we find that the AUC permutation importance and the accuracy permutation importance perform similarly, regardless of the outcome imbalance. Further, we did not find any consistent differences in terms of the features detected by either random forest or conditional inference forest even though conditional inference forest is supposed to be better suited for categorical data \cite{strobl_bias_2007}.

We also see this preference for features with smaller imbalances in the real data. For example, for school 1, we saw that the more balanced \textit{RaceLatinx} was detected over the less balanced \textit{RaceBlack} and \textit{RaceMulti} even though the theoretical odds ratio of \textit{RaceLatinx} was smaller than that of the other two features.

Across the real data and simulated data, we see the expected bias with the Gini importance in which the continuous features are ranked higher than any of the categorical features. This result is most noticeable in Fig.~\ref{fig:RF_bubble} plot A where only continuous features are detected and Fig.~\ref{fig:school1} plot B where all of the continuous noise features outrank all but one feature.

\textit{How might known machine learning biases manifest in traditionally explanatory techniques such as logistic regression?}
We see similar biases in logistic regression as we see in the random forest for feature selection. For a sample size of $N=100$ with a multiple comparison correction, we are unable to detect most features and even without correction, we can only detect low imbalance \textit{OR=3} features. The uncorrected results are similar to those of the permutation importances for the forest algorithms.

For $N=1000$, we can detect most \textit{OR=3} features and without correction, low imbalance \textit{OR=1.5} features. Again, the uncorrected logistic regression results resemble those of the forest algorithms but seem to be more aligned with the conditional inference forest results than the random forest results.

Once we get to a large sample size, $N=10,000$, we can detect nearly all features, just as we can for the forest algorithms. However, for logistic regression, we also get an occasional false positive. Given the size of the data, it is not unreasonable that the logistic regression model might be picking up on minor differences in the noise features which it treats as a signal.

With explanatory techniques like logistic regression, we could also investigate how well they estimated the built-in odds ratio. We found that while the built-in value is almost always in the confidence interval, this has more to do with the width of the intervals than the ability of the algorithms. In general, confidence interval width increases with feature and outcome imbalance and decreases with sample size. The decrease in width as the sample size increases corresponds to what we would expect based on the conclusions of \citeN{nemes_bias_2009}.

We also observed the same general trend for the real data. Features with higher imbalances tend to have the widest confidence intervals, which can span several orders of magnitude.

\textit{How might penalized regression techniques successfully applied in other disciplines be used in EDM to combat any discovered biases?}

While none of the five techniques we tried, Firth penalization, Log-F penalization, lasso, ridge, or elastic net, corrected the bias, they did show promise for use in future EDM studies.

For explanatory methods, Firth and Log-F were found to shrink the confidence intervals, especially for highly imbalanced features and highly imbalanced outcomes. While Firth can still show wide confidence intervals, the Firth confidence intervals were found to be smaller than those of traditional logistic regression. On the other hand, Log-F provided at worst similar performance to Firth penalization and for higher imbalances, seemed to shrink the width of the confidence interval more than Firth penalization did. We found that both of these methods were most useful for smaller data sets, $N=100$, while for the medium and larger data sets, their performance was similar.

When it came to the distributions of the estimated odds ratios, Log-F often showed a smaller distribution. While the results were comparable for the small data sets, for medium data sets and features with high imbalance, Firth penalization overestimated the odds ratio and had more variability. Conversely, Log-F produced more accurate and less variable distributions.

For predictive methods, lasso, ridge, and elastic net were only used in a bootstrap, so we cannot cannot discuss the confidence interval width. We can however discuss the distribution of estimated odds ratios.

For lasso and small data sets, we find that many of the features are shrunk to zero, especially for higher imbalances. For example, even for a small, balanced sample, many of the \textit{OR=1.5} features were shrunk to zero while the other methods did not treat them as consistent with noise. Elastic showed similar results although the effect was not as severe.

For ridge and small data sets, the distribution of the estimated odds ratio was often the smallest for a given data set. Given that ridge is designed to shrink the variability of the estimates, this finding is not surprising.

For medium data sets, lasso, ridge, and elastic net performed similarly to the other methods in terms of the distribution of estimated odds ratios.

While our results generally agree with other studies, a true comparison is difficult because each study used its own subset of the algorithms, including ones we used in our study as well as ones we did not. Therefore, which algorithm performed best and under what circumstances depends just as much on the algorithms it was compared to as the algorithm itself.

In general, other studies have tended to find that Firth penalization does outperform logistic regression in the case of outcome imbalance \cite{heinze_solution_2002,van_smeden_no_2016,kim_comparison_2014,doerken_penalized_2019} and Log-F penalization shows promise when working with imbalanced data and can outperform Firth-penalization \cite{ogundimu_prediction_2019,rahman_performance_2017}.

Likewise, studies like \citeN{pavlou_review_2016} have found that ridge penalization works well except when there are many noise features while lasso performs better when there are many noise features but limited correlations, which is consistent with our results. Their study also found that elastic seemed to perform well in all cases, which generally matches what we found.

In terms of our finding that none of the methods fixed the issues around feature or outcome imbalance, \citeN{van_calster_regression_2020} reported a similar finding for shrinkage techniques. Specifically, they found that despite working well on average, shrinkage techniques often did not work well on individual data sets, even in cases where the techniques could have provided the most benefit such as in small sample size or low events per variable cases. Even though the techniques did not solve any of the issues in our study, they still showed promise for reducing the scale of the confidence intervals and warrant greater adoption by the EDM communities.

\subsection{Limitations and Researcher Choices}
In this section, we shift our focus from the results of the research questions and instead consider how our choices around constructing the simulated data, tuning or not tuning our models, defining ``detected features,'' and assessing the models might have impacted the conclusions we can draw from this study.

\subsubsection{Our data sets}
For our simulation study, we used the same levels of information as in the \citeN{boulesteix_random_2012} study which we wished to extend. We followed their convention that $OR=3$ corresponded to a large effect while $OR=1.5$ corresponded to a moderate effect. However, \citeN{olivier_effect_2013} noted that what constitutes a large, medium, or small odds ratio depends on the feature imbalance, outcome imbalance, and correlations. Therefore, even though we are using the same odds ratios for the different imbalances, they might not necessarily contain the same amount of predictive or explanatory power in a ``large", ``medium", or ``small'' sense.

One noticeable difference between our study and the \citeN{boulesteix_random_2012} study was the number of features. In EDM studies, the number of features can vary from the order of 10 features \cite{mu_towards_2020,li_erroneous_2020,zhao_predicting_2020,nguyen_exploring_2020,bulathwela_predicting_2020} to the order of 100 features \cite{sanyal_feature_2020,aulck_increasing_2020,hur_harbingers_2020,bosch__2020,korosi_clickstream-based_2018} to over 1,000 features e.g. \cite{sanyal_feature_2020}. Furthermore, a recent review of 62 student performance prediction papers found that around a third of the papers had 3 or fewer features while over half had at least 4 features, with a maximum number of 254 features \cite{namoun_predicting_2021}. As our prior work has tended to have a number of features on the order of 10, a recent EDM simulation study also studied this number of features \cite{autenrieth_stacked_2021} and the Boulesteix study focused on data with the order of a hundred features, we chose to study data where the number of features is on the order of ten although other ranges are equally valid to study and should be investigated in future studies. 

Even at the order of 10 features however, one could argue that we still had too many features based on our sample size. For example, a rule of thumb is that there should be at least 10 cases of the minority outcome for each feature in the model, referred to as the events per variable \cite{peduzzi_simulation_1996,austin_events_2017}. In that case, we would have needed at least a sample size of 400 for the 50/50 outcome imbalance and a sample size of 2,000 for the 90/10 outcome imbalance case.

However, recent work has called into question whether this rule of thumb is supported by evidence \cite{van_smeden_no_2016}. \citeN{van_smeden_sample_2019} found that events per variable did not have a strong relation to predictive performance of models and instead, recommended that a combination of the number of predictors, the total sample size and the events fraction be used to assess sample size criteria. Likewise, \citeN{courvoisier_performance_2011} found that logistic regression can encounter problems even if the events per variable were greater than 10 and concluded that there is no single rule for guaranteeing an accurate estimate of parameters for logistic regression. Even if the rule of thumb were true for logistic regression, \citeN{pavlou_how_2015} claims that penalized regression is effective when the events per variable is less than 10.

\subsubsection{Hyperparameter tuning}
For our simulation study, we did not do extensive hyperparameter tuning for the forest algorithms. We did this because 1) \citeN{probst_hyperparameters_2019} found that random forest is robust against hyperparameter specification, its performance depends less on the hyperparameters than other machine learning methods, and its default choice of hyperparameters are often good enough and 2) \citeN{couronne_random_2018} states that for a method to become a standard tool (as random forest is in EDM), it needs to be easy to use by researchers without computational backgrounds and cannot involve complex human interaction, which is not true of hyperparameter tuning.

In addition, we only do hyperparameter tuning for lasso, ridge, and elastic because testing multiple values for $\lambda$ is built in to the \texttt{glmnet} algorithm that we used to run the models. Even then, recent work suggests that optimizing $\lambda$ for small or sparse data sets results in substantial variability of the coefficients and the found $\lambda$ might be negatively correlated with the optimal values, meaning that hyperparameter tuning might not have been advisable for our data in the first place \cite{sinkovec_tune_2021}.

However, for completeness and to minimize computation time needed, we did experiment with multiple choices for the number of trees in the forest, $n_{tree}$, and the number of features used for each tree, $mtry$. For the conditional inference forests with the AUC importance, a sample size of $N=1000$ and outcome imbalances of 50/50, 60/40, 70/30, 80/20, and 90/10, we tried $n_{tree}=\{50, 100, 500, 1000, 5000\}$ and $mtry=\{1, p/3, \sqrt{p},p/2, p\}$ where $p$ is the total number of features in the model. We did not find any meaningful differences in which features were selected and no set of the hyperparameters consistently performed better than the default ($n_{tree}=500, mtry=\sqrt{p}$). Therefore, we used the default choices for throughout the study.

\subsubsection{Determining Detected Features}
For the forest algorithms, we chose to use the simple and intuitive method of whether the feature ranked above the first noise feature to determine which features were detected. We were able to do this because we knew which features were noise and in the case of the real data, we added features we created to be noise. We could have, however, used a variety of other methods to detect features though each has its own limitations in the context of our study. See \citeN{hapfelmeier_new_2013} for an overview of different approaches, some comparisons, and their own novel method.

In general, the techniques for feature selection in forest algorithms fall into two broad categories. First, there are elimination techniques that pull out a subset of the features based on criteria. For example, \citeN{diaz-uriarte_gene_2006} used a recursive backward elimination technique that removes a certain fraction of features until only 2 remain. The technique then selects the model with the fewest features that performs within 1 standard error of the best model using whatever metric the researcher chooses. These type of methods are not appropriate for this study because they can restrict the features too much. That is, by having some cutoff or elimination procedure, features which contain only a small amount of predictive information could be eliminated even though they are in fact predictive.

The second common approach is to use some type of permutation test to generate a p-value. Under this approach, either the outcome or each individual feature is permuted and then run through the model to produce some metric. This is then done a large number of times to get a distribution of the metric. Then the unpermuted data is run through the model to get the actual value of the metric. The p-value is then the fraction of cases where the permuted metric is as extreme as the actual value of the metric \cite{ojala_permutation_2010}. This approach has been used in various random forest studies \cite{chen_forest-based_2007,wang_maximal_2010} and has been extended into the PIMP heuristic for correcting the Gini importance bias \cite{altmann_permutation_2010}. While these methods provide an analogous method for comparing with explanatory methods, they can be computationally intensive as they require the distributions to be conducted from scratch for each model.

As a way to reduce the computational complexity, \citeN{janitza_computationally_2016} proposed that the negative importances, which are assumed to be noise features because they are making the predictions worse, could be used to construct a null distribution. Under this approach, the distribution of the negative importances are reflected across the axis to create the distribution for positive values. The same procedure as above can then be used to calculate the p-values. While this procedure is computationally feasible, it may be of little use when the number of features is small and there are a limited number of features which could have negative importances.

\subsubsection{Assessing Our Models}
For our real data, we noticed that many of the models did not produce out-of-sample AUCs in the acceptable range of at least 0.7. Here we try to address that.

First, we acknowledge that one type of model should not always perform better than another; this is the basis of the ``no free lunch theorems'' for optimization \cite{wolpert_no_1997}. Various studies comparing logistic regression and random forest find a similar result where which algorithm performs best depends on the data set \cite{couronne_random_2018,kirasich_random_2018,walinder_evaluation_2014}. Therefore, the fact that logistic regression models perform better than the forest models is not necessarily a problem. In fact, by comparing multiple models and finding that some work better than others, we can have greater confidence that our results are detecting a signal in the data and not just modeling the random variations in the data.

Second, we need to acknowledge that overfitting is happening with the elastic net and conditional inference forests. This overfitting can be detected by looking for differences in the training and testing set AUCs, where a higher training AUC is characteristic of overfitting. The amount of overfitting seems worse for the smaller data sets as shown in Table \ref{tab:auc_table}. This result is not unexpected because with smaller data sets, there are fewer cases to learn from. The noise in the model might then be seen as a signal and treated as though it contains predictive information.

If we look at the other models and their results in Table \ref{tab:auc_table}, we notice that the forest models and elastic net perform best on the school 3 data sets, which correspond to the medium sized data sets in the simulation study and largest of the real data sets. For the forest algorithms specifically, they perform best on the school 3 shortlist data set, which happens to have a smaller outcome imbalance than school 3 admit. This result suggests that to effectively use the predictive approach, the data set should not be too imbalanced and based on the results for school 1 and school 2, the amount of data should be on the order of 1,000 cases.

Additionally, the higher AUC for logistic regression and Log-F might be thought of as their own type of overfitting. Due to the train/test procedure of the predictive paradigm, these two methods are working with the full data set rather than just 80\% of the cases, corresponding to a 25\% increase in data to work with and hence, learn from. With the ``extra'' data, these models might be better able to detect trends in the data and separate them out from noise.

While there do exist techniques for detecting overfitting in logistic regression, many of them use some type of testing or validation data set. For example, the Copas test of overfitting recommends splitting the data in half, using one half of the data to develop the regression model, using that model with the other half of the data to make predictions of the outcome, and then perform a linear regression with the predictions and actual values, testing whether the coefficient is different from 1 \cite{copas_regression_1983}. If it were, that would provide evidence of overfitting. However, this approach is nearly equivalent to using logistic regression in a predictive manner rather than in an explanatory manner.

For a technique that aligns the explantory nature of logistic regression, we can examine the residual plots. Because, logistic regression produces discrete residuals, using binned residual plots instead might be helpful \cite{gelman_data_2006}. Under this approach, cases are divided into bins and the average value in each bin is plotted against the average residual in that bin. This approach allows the otherwise binary residual to take on any value of the form $\frac{i}{n_{bin}}$ where $n_{bin}$ is the number of cases in the bin and $\{ i \in \mathbb{Z} : -n_{bin} \leq i \leq n_{bin} \}$.

When implemented via the \texttt{arm} package \cite{gelman_arm_2020}, 95\% confidence intervals are generated and we can get an idea of how good the model is by examining what fraction of the binned residuals fall within the intervals. When we do so, we find that the fraction of residuals falling outside of the confidence intervals are between 0.20 for School 3 shortlist and 0.34 for School 3 admit, suggesting the models might in fact, not fit well. There does not appear to be a pattern based on the sample size or outcome imbalance. The plots are shown in Fig.~\ref{fig:residual_plot} in the appendix.

\section{Future Work}\label{sec:future_work}
While we considered six approaches to logistic regression, two machine learning algorithms, and three importance measures, these are not the only approaches we could have used. Indeed, these are not even the only logistic regression or random forest techniques we could have used but chose these algorithms as a starting point. Future work could then consider how other modifications of logistic regression or random forest might improve upon the problems we have identified here.

For example, for logistic regression algorithms, \citeN{puhr_firths_2017} proposed two modifications to Firth penalization, a post-hoc adjustment of the intercept and iterative data augmentation, that showed promise in their simulation study. Based on their results, they recommend using their methods or penalization by Cauchy Priors \cite{gelman_weakly_2008}, which we did not include in this study, as better options than Log-F when confidence intervals were of interest. Furthermore, a later study comparing logistic regression, Firth penalization, and the modifications to Firth penalization found that the modifications to Firth's method worked best in terms of parameter estimation bias for rare events and small sample cases \cite{olmus_comparison_2019}.

In terms of forest algorithms, there are several variants that might be useful for the data we encounter in EDM studies. For example, Balanced Random Forest and Weighted Random Forest have been developed for working with imbalanced outcomes \cite{chen_using_2004} and Oblique Random Forests have been developed to allow for diagonal cuts in the feature space rather than the horizontal or vertical cuts allowed under traditional random forest algorithms \cite{menze_oblique_2011}. In their study, \citeN{menze_oblique_2011} found that Oblique Random Forests outperform traditional random forest when the data is numerical rather than discrete, which might show promise for our data depending on the ratio of numerical features to categorical or binary features.

Alternatively, there are non-CART-based approaches to random forest \cite{breiman_classification_1984}. \citeN{loh_variable_2021} conducted a simulation study of various approaches to random forest and variable importance, finding that forests grown using the GUIDE algorithm, which is implemented for both classification \cite{loh_improving_2009} and regression \cite{loh_regression_2002}, was unbiased while the random forest and conditional inference forest approaches we used here were not. In their study, a method was unbiased ``if the expected values of its scores are equal when all variables are independent of the response variable,'' which would correspond to a case in our study where the odds ratios were 1 for all features. Nevertheless, such an approach might still be worth looking into.

There are also newer importance measures that show promise. In a simulation study, \citeN{nembrini_revival_2018} proposed a modification of the Gini importance, which they claim removed its bias toward features with more categories and the biases observed here regarding feature imbalance. However, their simulated studies with feature importance only considered null cases in which none of the features were predictive of an outcome. Nevertheless, further study of this approach might be fruitful.

In contrast to the algorithms used to analyze the data, future work should also explore how changes to the data itself might affect algorithm performance. For example, we could use the risk ratio to encode the level of information in a feature instead of the odds ratio. In theory, risk ratio provides a more intuitive way to quantify the amount of information in a feature because it is based on the ratio of probabilities rather than a ratio of odds.  \citeN{zhang_whats_1998} proposed a method to convert the odds ratio to a risk ratio, though more recent work has called this approach into question and suggests alternatives \cite{mcnutt_estimating_2003,karp_re_2014}. Because these two measures are related but not the same, there might be additional insights related to which features are detected based on how we define the amount of ``predictiveness'' they have.

To better replicate real EDM data, future work could also explore how the amount of correlation between the features affects the results. In the case of correlated features, new issues with permutations emerge, including the model needing to extrapolate to regions where the model was not trained to calculate a feature importance \cite{hooker_please_2019}. Various approaches have been developed for correlated data with forest algorithms \cite{strobl_conditional_2008,hooker_please_2019,molnar_model-agnostic_2020}, which warrant future study, especially for the type of data we see in EDM studies.

Additionally, future work can extend the data beyond binary features and include categorical features. While logistic regression often requires categorical features to be binarized, binarizing categorical features into a single new feature can cause a loss of information. For example, treating exam responses as correct or incorrect hides information about the specific incorrect answer the student chose and possible patterns \cite{springuel_reconsidering_2019} and combining demographics into a single ``underrepresented'' category can hide the struggles of students of different races and ethnicities \cite{shafer_impact_2021}. As random forest implementations can often handle categorical features directly and logistic regression can handle categorical features through one-hot encoding, future work can consider how the biases explored in this study might manifest in categorical data and how aggregating or segregating features might introduce its own biases. Ordinal categorical data should be of special interest given that one-hot encoding results in the loss of the ordinal information.

Finally, future work can consider how both the data and algorithms affect how features are detected. Recently, \citeN{pangastuti_data_2021} found that a combination of bagging, boosting, and SMOTE improved random forest's classification ability on a large, imbalanced educational data set. We need to be careful with such approaches, however, because our data is not just data but represents actual students. Therefore, we need to be certain that our conclusions are based on the student data and not simulated students ``created'' to make the data easier to analyze.

\section{Conclusion and Recommendations}\label{sec:conclusion}
Our work suggests that for both predictive and explanatory models, feature and outcome imbalances can cause algorithms to detect different features despite the same built-in amount of information. We found this to be true for random forest, conditional inference forest, logistic regression, as well as in various penalized regression algorithms. On a practical level, this means that if we are using these algorithms for determining which features might be related to some outcome of interest, we might be introducing false negatives into our results, potentially missing factors that are related to the outcome.

Based on the results of this study, we propose three recommendations for EDM researchers. First, for smaller data sets with highly imbalanced features, we recommend using a penalized version of logistic regression such as Log-F. Even though Firth penalization was often comparable to Log-F, Firth penalization is not implemented in all statistical software. Log-F however can be used with any statistical software that can perform logistic regression because it is based on data augmentation. If the outcome is also imbalanced, it is even more essential to consider penalized approaches.

Second, for medium or large data sets ($N \geq 1,000$) traditional logistic regression and random forest or conditional inference forest with a permutation importance perform similarly, so either approach works. While the algorithms still do not perform perfectly, none of them provided a consistent advantage over another. We recommend that researchers first consider whether the research questions are best answered using predictive or explanatory techniques and then which affordances of the algorithms are most relevant to the study.

Finally, we call on researchers to include information about their features in their publications, including the features themselves, their distributions, and in the case of categorical or binary features, their class frequencies as others outside the EDM communities have done in the past \cite{greenland_sparse_2016}. A simple example of how this might be done is shown in Table \ref{tab:igen_imbalance}. In addition, we recommend that researchers include data set characteristics or so-called ``meta-features'' as well. Some examples include sample size, the number of features, the number of numerical features, the number of categorical features, and the percentage of observation of the majority class or outcome balance \cite{couronne_random_2018}. Just as there have been calls for increased reporting of demographics in the EDM community and outside of it to understand how results might depend on the sample population or generalize \cite{paquette_whos_2020,kanim_demographics_2020}, we are calling for the same with the explanatory and predictive models we create, partially addressing some of the questions raised by \citeN{knaub_two-phase_2019} in their analysis of physics education research quantitative work. By doing so, we hope for greater acknowledgement of possible sources of bias or false negatives in feature selection as a result of the data or algorithms used in EDM studies.


\section{Acknowledgments}
We would like to thank Julie Posselt, Casey Miller, and the Inclusive Graduate Network for providing us with the graduate admissions data sets for this project. We would also like to thank Rachel Henderson, Morten Hjorth-Jensen, and Arjun Krishnan for their feedback and suggestions in the early stages of this project. This project was supported by the Michigan State University College of Natural Sciences and the Lappan-Phillips Foundation.

\bibliographystyle{acmtrans}
\bibliography{references.bib}

\section{APPENDIX}
Here we provide the plots for the confidence intervals and percentile bootstraps for the $N=1000$ data set and the additional three data sets from Sec.~\ref{sec:realdata} for completeness. The plots show the same general results as discussed in the main manuscript. We also include the residual plots from the discussion.

\begin{figure}[hbt!]
    \centering
    \includegraphics[width=1\linewidth]{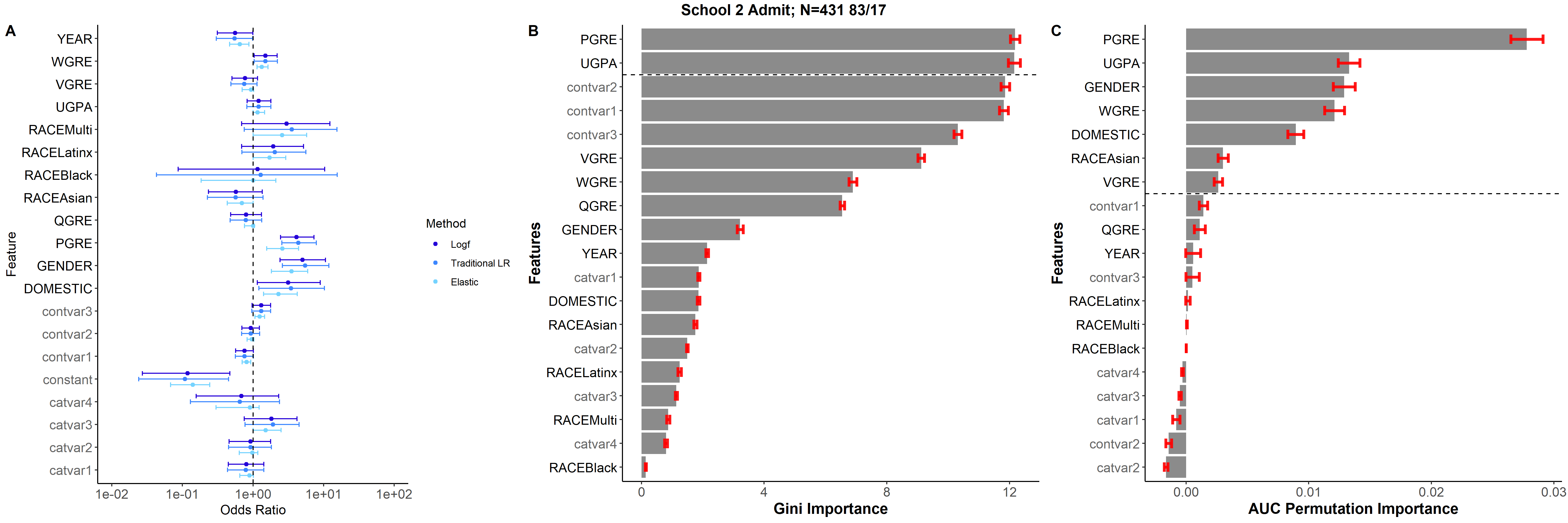}
    \caption{Comparison of the odds ratio, Gini importance, and AUC-permutation importance for the features in the school 2 admit data set.}
    \label{fig:school2}
\end{figure}

\begin{figure}
    \centering
    \includegraphics[width=1\linewidth]{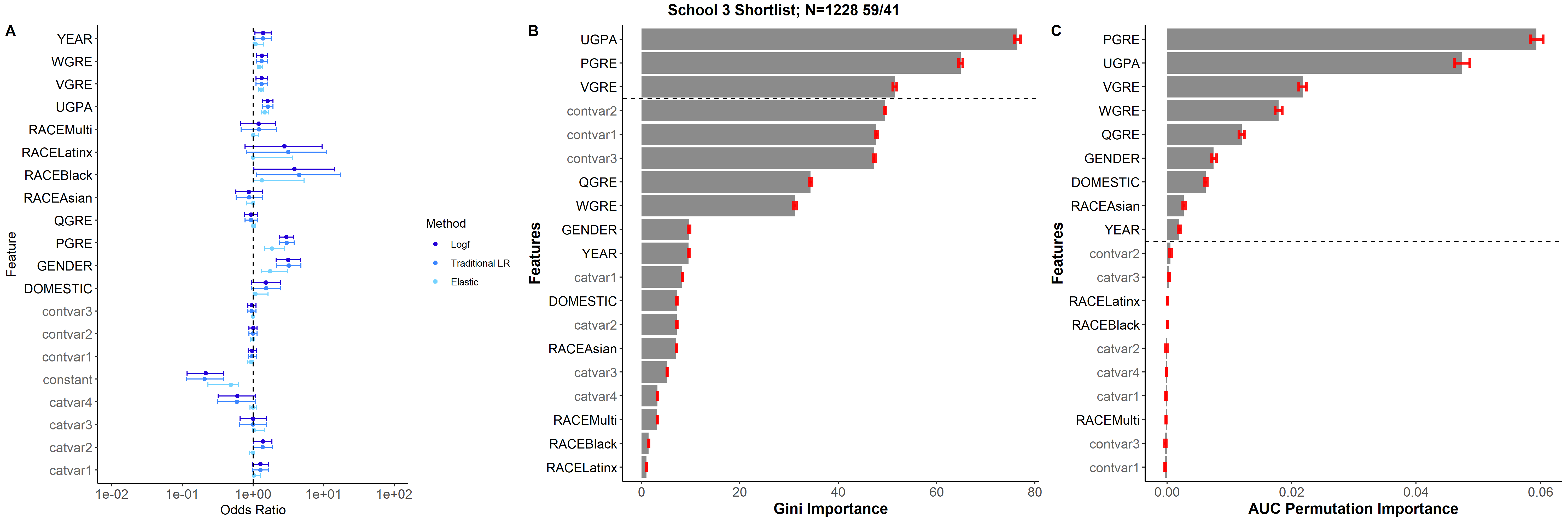}
    \caption{Comparison of the odds ratio, Gini importance, and AUC-permutation importance for the features in the school 3 shortlist data set. }
    \label{fig:school3_shortlist}
\end{figure}

\begin{figure}
    \centering
    \includegraphics[width=1\linewidth]{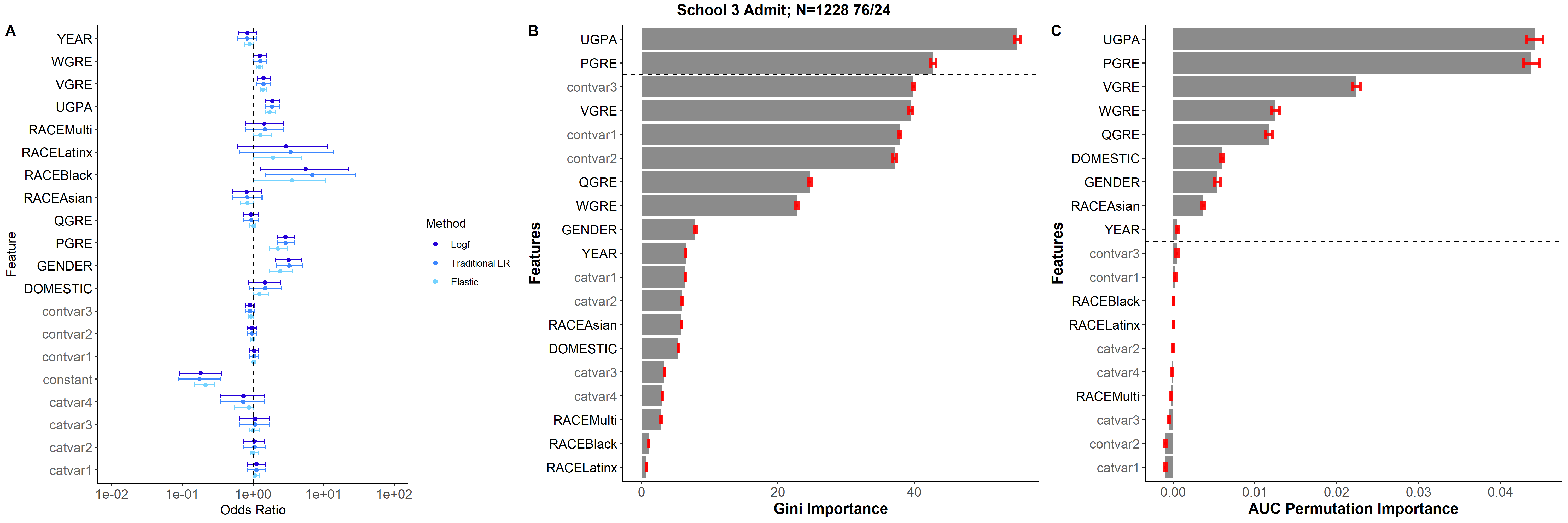}
    \caption{Comparison of the odds ratio, Gini importance, and AUC-permutation importance for the features in the school 3 admit data set. }
    \label{fig:school3_admit}
\end{figure}

\begin{figure}
    \centering
    \includegraphics[width=0.8\linewidth]{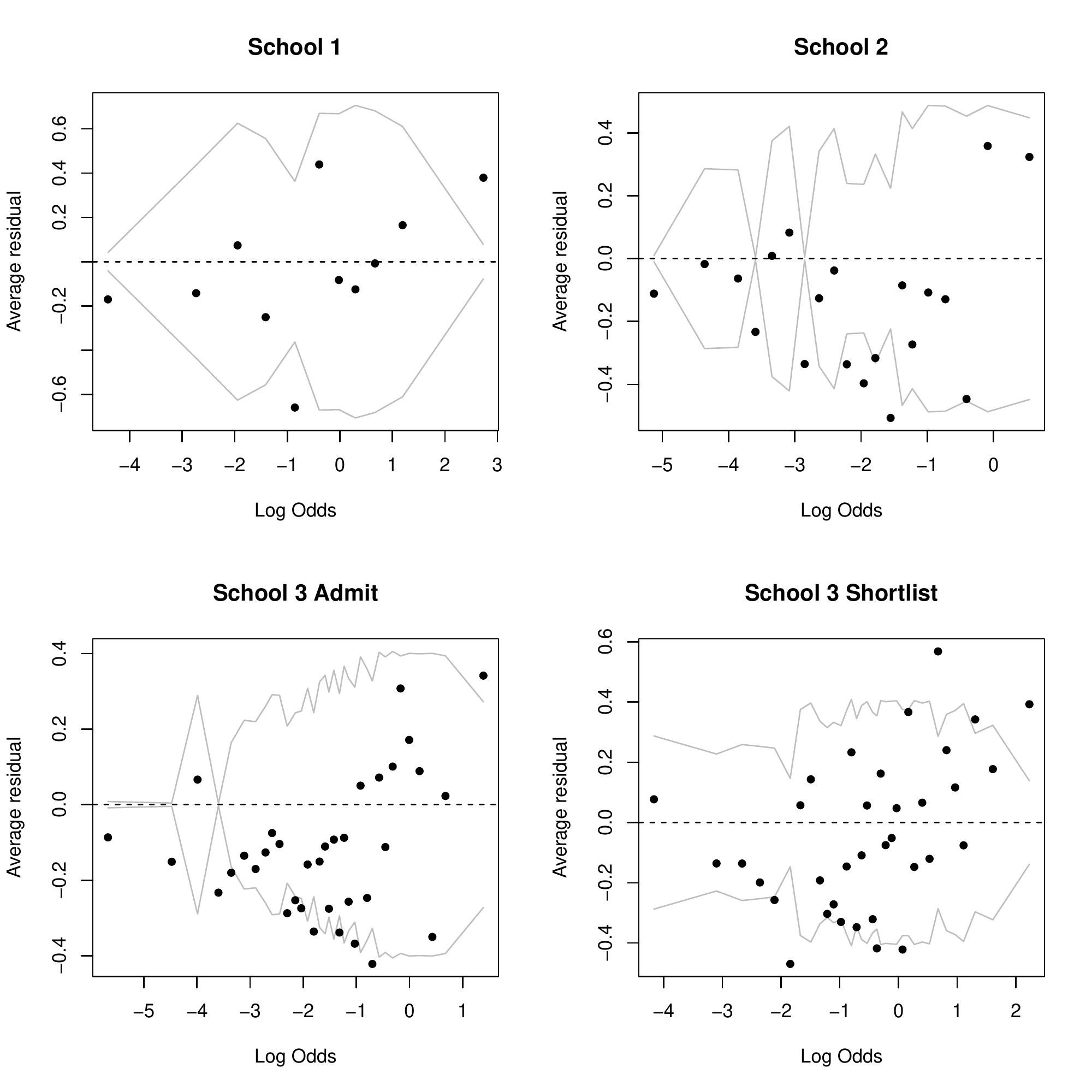}
    \caption{Plots of the Log-odds vs the average residual in each bin for the four schools. Across all plots, between 20\% and 34\% of the points fall outside of the confidence intervals, suggesting the logistic regression models might not be fitting the data especially well.}
    \label{fig:residual_plot}
\end{figure}

\end{document}